\documentclass{article}%
\usepackage{amsmath}
\usepackage{amsfonts}
\usepackage{amssymb}
\usepackage{graphicx}%
\newtheorem{theorem}{Theorem}

\newtheorem{definition}[theorem]{Definition}

\newtheorem{proposition}[theorem]{Proposition}
\newtheorem{remark}[theorem]{Remark}

\newcommand{\bpartial}{\mathop{\partial\kern -4pt\raisebox{.8pt}{$|$}}}
\newcommand{\bra}{\mathopen{[\kern-1.6pt[}}
\newcommand{\ket}{\mathclose{]\kern-1.5pt]}}
\newcommand{\bbra}{\mathopen{[\kern-2.2pt[\kern-2.3pt[}}
\newcommand{\bket}{\mathclose{]\kern-2.1pt]\kern-2.3pt]}}

\makeindex
\begin{document}

\title { Classification of real low dimensional Jacobi (generalized)-Lie bialgebras}
\vspace{3mm}

\author {  A. Rezaei-Aghdam \hspace{-2mm}{ \footnote{ rezaei-a@azaruniv.edu }} \hspace{2mm}{\small and}\hspace{2mm}
 M. Sephid{ \footnote{s.sephid@azaruniv.edu }}\hspace{2mm}\\
{\small{\em Department of Physics, Faculty of Sciences, Azarbaijan Shahid
}}\\
{\small{\em Madani University  , 53714-161, Tabriz, Iran  }}}

\maketitle

\vspace{3cm}

\begin{abstract}

We describe the definition of Jacobi (generalized)-Lie bialgebras $(({\bf{g}},\phi_{0}),({\bf{g}}^{*},X_{0}))$ in terms of structure constants of the Lie algebras ${\bf{g}}$ and ${\bf{g}}^{*}$ and components of their 1-cocycles $X_{0}\in {\bf{g}}$ and $\phi_{0}\in {\bf{g}}^{*}$ in the basis of the Lie algebras. Then, using adjoint representations and automorphism Lie groups of Lie algebras, we give a method for classification of real low dimensional Jacobi-Lie bialgebras. In this way, we obtain and classify real two and three dimensional Jacobi-Lie bialgebras.
\end{abstract}

\newpage
%%%%%%%%%%%%%%%%%%%%%%%%%%%%%%%%%%%%%%%%%%%%%%%%%%%%%%%%%%%%%%%%%%%%%%%%%%%%%%%%%%%%%%%%%%%%%%%%%%
\section{\bf Introduction}

Jacobi structure on a manifold $M$ (Jacobi manifold) has been introduced by A. Lichnerowicz \cite{Lich2} and, as a local Lie algebra structure on $C^{\infty}(M,R)$,
by A. Kirillov \cite{Ki}. Jacobi structures have all properties of the Poisson structures, except that they are not necessarily derivation. Generalization of Poisson-Lie groups (a Lie group whose Poisson structure, is compatible with the group structure  \cite{Drin},\cite{Drin2}) to Jacobi-Lie groups has been done in
\cite{Iglesias}. Of course, first, this work has been done for the Lie bialgebroids, namely, the relation between Jacobi structure and Lie bialgebroids has been studied, in addition, the generalized Lie bialgebroids \cite{IM} (Jacobi bialgebroid \cite{GM}) have been defined; then the generalized Lie bialgebras and their group structure (i.e., Jacobi-Lie groups) have been defined \cite{Iglesias}. Generalized Lie bialgebras introduced by D. Iglesias and J. C. Marrero are the algebraic structures of Jacobi-Lie groups, similar to Lie bialgebras as the algebraic structures of the Poisson-Lie groups \cite{Drin},\cite{Drin2}. In \cite{Iglesias}, a generalizing Yang-Baxter equation method has been proposed to obtain Jacobi-Lie bialgebras and some examples of Jacobi-Lie bialgebras have been given. Here, we describe the definition of the Jacobi-Lie bialgebras  $(({\bf{g}},\phi_{0}),({\bf{g}}^{*},X_{0}))$ in terms of structure constants of the Lie algebras ${\bf{g}}$ and ${\bf{g}}^{*}$ and components of their 1-cocycles $X_{0}\in {\bf{g}}$ and $\phi_{0}\in {\bf{g}}^{*}$ in the basis of the Lie algebras. Then, using adjoint representations and automorphism Lie groups of these Lie algebras, we obtain a method for classifying Jacobi-Lie bialgebras and classify real two and three dimensional Jacobi-Lie bialgebras. The outline of the paper is as follows.

In section two, we describe the definition of the Jacobi-Lie bialgebras $(({\bf{g}},\phi_{0}),({\bf{g}}^{*},X_{0}))$ in terms of the structure constants of the Lie algebras ${\bf{g}}$ and ${\bf{g}}^{*}$ and components of 1-cocycles $X_{0}\in {\bf{g}}$ and $\phi_{0}\in {\bf{g}}^{*}$ in the basis of the Lie
algebras; and at the end of this section we give a proposition about equivalence of Jacobi-Lie bialgebras using automorphism of Lie algebras. Then, in section
three we give the matrix representation of the obtained relations (in section two) using adjoint representations. Also, we give three steps for obtaining and
classifying real low dimensional Jacobi-Lie bialgebras. In section four, in order to clarify our method, we give a detailed example, then we obtain and classify real two and three dimensional Jacobi-Lie bialgebras by this method. Some remarks are addressed in conclusion.
%%%%%%%%%%%%%%%%%%%%%%%%%%%%%%%%%%%%%%%%%%%%%%%%%%%%%%%%%%%%%%%%%%%%%%%%%%%%%%%%%%%%%%%%%%%%%%%%%%

\section{\bf Jacobi-Lie bialgebra }
In this section, we review the basic definitions of Jacobi (generalized)-Lie bialgebras $(({\bf{g}},\phi_{0}),({\bf{g}}^{*},X_{0}))$. Then, we describe these definitions in terms of structure constants of the Lie algebras ${\bf g}$ and ${{\bf{g}}^{*}}$ and components of 1-cocycles $X_{0}\in {\bf{g}}$ and $\phi_{0}\in {\bf{g}}^{*}$ in the basis of the Lie algebras.

\begin{definition}\cite{Iglesias}: A Jacobi-Lie bialgebra is a pair $(({\bf{g}},\phi_{0}),({\bf{g}}^{*},X_{0}))$,
where $({\bf{g}},[,]^{{\bf{g}}})$ is a real Lie algebra of finite dimension with Lie bracket $[,]^{{\bf{g}}}$, so that the dual space ${\bf{g}}^{*}$ is also a Lie algebra with bracket $[,]^{{\bf{g^{*}}}}$, $X_{0}\in {\bf{g}}$ and $\phi_{0}\in {\bf{g}}^{*}$ are 1-cocycles on ${\bf{g}}^{*}$ and ${\bf{g}}$,
respectively, and  {\small $\forall X,Y \in {\bf{g}}$} we have
\begin{equation}\label{1}
d_{*X_{0}}[X,Y]^{{\bf{g}}}=[X,d_{*X_{0}}Y]^{{\bf{g}}}_{\phi_{0}}-[Y,d_{*X_{0}}X]^{{\bf{g}}}_{\phi_{0}},
\end{equation}
\begin{equation}\label{2}
\phi_{0}(X_{0})=0,
\end{equation}
\begin{equation}\label{3}
i_{\phi_{0}}(d_{*}X)+[X_{0},X]=0,
\end{equation}
\end{definition}
where $i_{\phi_{0}}P$ is contraction of a $P\in\wedge^{k}{\bf g}$ to a tensor $\wedge^{k-1}{\bf g}$; furthermore $d_{*}$ being the Chevalley-Eilenberg differential of ${\bf{g}}^{*}$ acting on ${\bf{g}}$ and  $d_{*X_{0}}$ is its generalization such that we have
\begin{equation}\label{4}
d_{*X_{0}}Y= d_{*}Y+X_{0}\wedge Y,
\end{equation}
meanwhile $[,]^{{\bf{g}}}_{\phi_{0}}$ is $\phi_{0}$-Schouten-Nijenhuis bracket with the following properties
$$
\hspace{-11cm}\forall P\in\wedge^{k}{\bf{g}}, P{'}\in\wedge^{k{'}}{\bf{g}}, P^{''}\in\wedge^{k^{''}}{\bf{g}},~~~~~~
$$
\vspace{-2mm}
\begin{equation}\label{5}
[P,P{'}]_{\phi_{0}}=[P,P{'}]+(-1)^{k+1}(k-1)P\wedge i_{\phi_{0}}P{'}-(k{'}-1)i_{\phi_{0}}P\wedge P{'},
\end{equation}

\begin{equation}\label{6}
[P,P{'}]_{\phi_{0}}=(-1)^{kk{'}}[P{'},P]_{\phi_{0}},
\end{equation}

\begin{equation}\label{7}
[P,P{'}\wedge P^{''}]_{\phi_{0}}=[P,P{'}]_{\phi_{0}}\wedge P^{''} + (-1)^{k{'}(k+1)}P{'}\wedge[P,P^{''}]_{\phi_{0}}-(i_{\phi_{0}}P)\wedge P^{'}\wedge P^{''},
\end{equation}

\begin{equation}\label{8}
(-1)^{kk^{''}}[[P,P{'}]_{\phi_{0}},P^{''}]_{\phi_{0}}+(-1)^{k{'}k^{''}}[[P^{''},P]_{\phi_{0}},P{'}]_{\phi_{0}}+
(-1)^{kk{'}}[[P{'},P^{''}]_{\phi_{0}},P]_{\phi_{0}}=0.
\end{equation}
Moreover, in the above definition, the $\phi_{0}(X_{0})$ means the natural inner product of the dual spaces ${\bf{g}}$ and ${\bf{g}}^{*}$. Note that, the above definition is symmetric with respect to $({\bf{g}},\phi_{0})$ and $({\bf{g}}^{*},X_{0})$ i.e., if $(({\bf{g}},\phi_{0}),({\bf{g}}^{*},X_{0}))$ is a Jacobi-Lie bialgebra then $(({\bf{g}}^{*},X_{0}),({\bf{g}},\phi_{0}))$ is also a Jacobi-Lie bialgebra, for this case, we have $d$ as the Chevalley-Eilenberg differential of ${\bf{g}}$ acting on ${\bf{g}}^{*}$ and it has the following $\phi_{0}\in{\bf{g}}^{*}$ generalization
\begin{equation}\label{9}
\forall w\in \wedge^{k} {\bf g^{*}} ~~~~~~~~~~~~~~~ d_{\phi_{0}}w=dw+\phi_{0}\wedge w,
\end{equation}
with the Schouten-Nijenhuis bracket replaced by

\begin{equation}\label{10}
[Q,Q{'}]^{\bf g^{*}}_{X_{0}}=[Q,Q{'}]^{\bf g^{*}}+(-1)^{k+1}(k-1)Q\wedge i_{X_{0}}Q{'}-(k{'}-1)i_{X_{0}}Q\wedge Q{'},
\end{equation}
$\forall~Q\in\wedge^{k}{\bf g^{*}}$,$Q'\in\wedge^{k'}{\bf g^{*}}$; with the properties \eqref{6}-\eqref{8} similar to $[ , ]^{{\bf{g}}}_{\phi_{0}}$.
\begin{remark}\cite{Iglesias}: In the above definition, $X_{0}$ and $\phi_{0}$ are 1-cocycles on ${\bf{g}}^{*}$ and ${\bf{g}}$, respectively, i.e., we must have

\begin{equation}\label{11}
d_{*}X_{0}=0,
\end{equation}
\begin{equation}\label{12}
d\phi_{0}=0.
\end{equation}
\end{remark}

\begin{remark}: In the case of $\phi_{0}=0$ and $X_{0}=0$ the definition 1 recovers the concept of a Lie bialgebra \cite{Drin}, that is, a pair of dual Lie algebras $({\bf{g}},{\bf{g}}^{*})$ such that relation \eqref{1} reduces to the following one

\begin{equation}\label{13}
d_{*}[X,Y]^{{\bf{g}}}=[X,d_{*}Y]^{{\bf{g}}}-[Y,d_{*}X]^{{\bf{g}}}.
\end{equation}
\end{remark}
For the above case, there is a correspondence between Lie bialgebra $({\bf{g}},{\bf{g}}^{*})$ and the Manin triple $({\bf{g}}\oplus {\bf{g}}^{*}, {\bf{g}},{\bf{g}}^{*})$ such that the direct sum ${\bf{g}}\oplus {\bf{g}}^{*}$  is a Lie algebra when ${\bf{g}}$ and ${\bf{g}}^{*}$ are isotropic subspaces of ${\bf{g}}\oplus {\bf{g}}^{*}$ with respect to $ad$-invariant symmetric pairing \cite{Drin2}. But, for the Jacobi-Lie bialgebra  $(({\bf{g}},\phi_{0}),({\bf{g}}^{*},X_{0}))$ in the sense of Tan and Liu \cite{TL}, we have the following bilinear skew-symmetric bracket on the space ${\bf{g}}\oplus {\bf{g}}^{*}$\\

~~~~$[X\oplus \zeta,Y\oplus \eta]^{{\bf{g}}\oplus {\bf{g}}^{*}}=([X,Y]^{{\bf{g}}}+({\cal{L}}_{*X_{0}})_{\zeta}Y-({\cal{L}}_{*X_{0}})_{\eta}X-\frac{1}{2}(\zeta(Y)-\eta(X))X_{0})$\\
\begin{equation}\label{14}
\oplus ([\zeta,\eta]^{{\bf{g}}^{*}}+({\cal{L}}_{\phi_{0}})_{X}\eta-({\cal{L}}_{\phi_{0}})_{Y}\zeta+\frac{1}{2}(\zeta(Y)-\eta(X))\phi_{0}),
\end{equation}
\vspace{2mm}
$\forall X,Y \in {\bf{g}}$ and $\zeta,\eta \in {\bf{g}}^{*}$; such that Lie derivative ${\cal{L}}_{*X_{0}}$ (resp. ${\cal{L}}_{\phi_{0}}$) of ${\bf{g}^{*}}$(resp. ${\bf{g}}$) on ${\bf{g}}$(resp. ${\bf{g^{*}}}$) are defined as follows{\footnote{For general definition of the differential and the Lie derivative associated with a 1-cocycle, one can see \cite{Iglesias,IM,LLMP}.}

\begin{equation}\label{15}
\hspace{-1.4cm}\forall w\in \wedge^{k} {\bf g^{*}} , X\in{\bf g} ~~~~~~~ ({\cal{L}}_{\phi_{0}})_{X} w = (d_{\phi_{0}} \circ i_{X}+i_{X} \circ d_{\phi_{0}}) w,
\end{equation}
and
\begin{equation}\label{17}
\forall~\xi\in{\bf g^{*}} ~~~~~~~~~~ ({\cal{L}}_{*X_{0}})_\xi {P} = (d_{*X_{0}} \circ i_{\xi}+i_{\xi} \circ d_{*X_{0}}) P,
\end{equation}
where, for  $P\in \wedge^{k} {\bf g}$ (resp. $w\in \wedge^{k} {\bf g^{*}}$),~$\phi_{0}$ (resp. $X_{0}$)-Lie derivative are defined by $\phi_{0}$ (resp. $X_{0}$)-Schouten Nijenhuis brackets in the following forms

\begin{equation}\label{16}
({\cal{L}}_{\phi_{0}})_{X} P = [X,P]^{{\bf{g}}}_{\phi_{0}},
\end{equation}
and
\begin{equation}\label{18}
({\cal{L}}_{*X_{0}})_{\xi} w = [\xi,w]^{{\bf{g}}^{*}}_{X_{0}}.
\end{equation}
In general, the $({\bf{g}}\oplus {\bf{g}}^{*}, [,]^{{\bf{g}}\oplus {\bf{g}}^{*}})$ is not a Lie algebra, i.e., the Jacobi identities do not satisfy the algebra ${\bf{g}}\oplus {\bf{g}}^{*}$ \cite{Iglesias,TL}. Now, choosing the basis of the Lie algebras ${\bf{g}}$ and ${\bf{g}}^{*}$ as $\{X_{i}\}$ and $\{\tilde{X}^{i}\}$, respectively, we try to express the above definitions in terms of structure constants. We have

\begin{equation}\label{19}
[X_{i},X_{j}]={f_{ij}\hspace{0cm}}^{k} X_{k}\hspace{1mm} , \hspace{1mm}[\tilde{X}^{i},\tilde{X}^{j}]={{\tilde{f}}^{ij}\hspace{0cm}}_{k} {\tilde{X}}^{k},
\end{equation}
where ${f_{ij}\hspace{0cm}}^{k}$ and ${{\tilde{f}}^{ij}\hspace{0cm}}_{k}$ are the structure constants of the Lie algebras ${\bf{g}}$ and ${\bf{g}}^{*}$, respectively, such that they satisfy the following Jacobi identities

\begin{equation}\label{20}
{f}_{ij}\hspace{0cm}^k{{f}_{km}}\hspace{0cm}^{n}+
{f}_{ik}\hspace{0cm}^n{{f}_{mj}}\hspace{0cm}^{k} +
{f}_{jk}\hspace{0cm}^n{{f}_{im}}\hspace{0cm}^{k}=0,
\end{equation}

\begin{equation}\label{21}
{\tilde{f}}^{ij}\hspace{0cm}_k{\tilde{f}^{km}}\hspace{0cm}_{n}+
{\tilde{f}}^{im}\hspace{0cm}_k{\tilde{f}^{jk}}\hspace{0cm}_{n} +
{\tilde{f}}^{jm}\hspace{0cm}_k{\tilde{f}^{ki}}\hspace{0cm}_{n}=0.
\end{equation}
Furthermore, according to duality between ${\bf{g}}$ and ${\bf{g}}^{*}$ we have

\begin{equation}\label{22}
<X_{i},\tilde{X}^{j}> = {\delta_{i}}\hspace{0cm}^{j}.
\end{equation}
On the other hand, we know that for the Lie bialgebras by choosing \cite{RHR}

\begin{equation}\label{23}
d_{*}X_{i}=-\frac{1}{2} {\tilde{f}^{jk}}\hspace{0cm}_{i} X_{j}\wedge X_{k},
\end{equation}
the relation \eqref{13} can be rewritten in terms of ${f_{ij}\hspace{0cm}}^{k}$ and
${{\tilde{f}}^{ij}\hspace{0cm}}_{k}$ as the following mixed-Jacobi identities

\begin{equation}\label{24}
{f}_{ij}\hspace{0cm}^k{\tilde{f}^{mn}}\hspace{0cm}_{k}=
{f}_{ik}\hspace{0cm}^m{\tilde{f}^{kn}}\hspace{0cm}_{j} +
{f}_{ik}\hspace{0cm}^n{\tilde{f}^{mk}}\hspace{0cm}_{j}+
{f}_{kj}\hspace{0cm}^m{\tilde{f}^{kn}}\hspace{0cm}_{i}+
{f}_{kj}\hspace{0cm}^n{\tilde{f}^{mk}}\hspace{0cm}_{i},
\end{equation}
and using the $ad$-invariant symmetric bilinear form on the Manin triple of Lie bialgebras ${\bf{g}}\oplus {\bf{g}}^{*}$,  one can find the following commutation relation \cite{Drin}

\begin{equation}\label{25}
[X_i , \tilde{X}^j] ={\tilde{f}^{jk}}\hspace{0cm}_{i} X_k +{f}_{ki}\hspace{0cm}^{j} \tilde{X}^k,
\end{equation}
where the relation \eqref{24} together with \eqref{20} and \eqref{21} are the Jacobi identities on the Lie algebra ${\bf{g}} \oplus {\bf{g}}^{*}$. Now, for Jacobi-Lie bialgebra \eqref{1}-\eqref{3} and \eqref{11}-\eqref{12}  one can also apply the relation \eqref{23} as Chevalley-Eilenberg differential. In this way, expanding $X_{0}\in {\bf{g}}$ and $\phi_{0}\in {\bf{g}}^{*}$ in terms of the basis of the Lie algebras ${\bf{g}}$ and ${\bf{g}}^{*}$

\begin{equation}\label{26}
X_{0}={\alpha}^{i} X_{i}~~~~~~~,~~~~~~~~\phi_{0}={\beta}_{j}{\tilde{X}}^{j},
\end{equation}
and using \eqref{4}, \eqref{19} and \eqref{23}, after some calculations, the relations \eqref{1}-\eqref{3} and \eqref{11}-\eqref{12} can be rewritten as follows, respectively,

$$
\hspace{-.3cm}{f}_{ij}\hspace{0cm}^k{\tilde{f}^{mn}}\hspace{0cm}_{k}-{f}_{ik}\hspace{0cm}^m{\tilde{f}^{kn}}\hspace{0cm}_{j} -\\
{f}_{ik}\hspace{0cm}^n{\tilde{f}^{mk}}\hspace{0cm}_{j}-{f}_{kj}\hspace{0cm}^m{\tilde{f}^{kn}}\hspace{0cm}_{i}-\\
{f}_{kj}\hspace{0cm}^n{\tilde{f}^{mk}}\hspace{0cm}_{i}+\beta_{i}{\tilde{f}^{mn}}\hspace{0cm}_{j}-\beta_{j}{\tilde{f}^{mn}}\hspace{0cm}_{i}+\alpha^{m}{f}_{ij}\hspace{0cm}^n-\alpha^{n}{f}_{ij}\hspace{0cm}^m\\
$$
\begin{equation}\label{27}
+(\alpha^{k}{f}_{ik}\hspace{0cm}^m-\alpha^{m}\beta_{i})\delta_{j}\hspace{0cm}^{n}
-(\alpha^{k}{f}_{jk}\hspace{0cm}^m-\alpha^{m}\beta_{j})\delta_{i}\hspace{0cm}^{n}-(\alpha^{k}{f}_{ik}\hspace{0cm}^n-\alpha^{n}\beta_{i})\delta_{j}\hspace{0cm}^{m}
+(\alpha^{k}{f}_{jk}\hspace{0cm}^n-\alpha^{n}\beta_{j})\delta_{i}\hspace{0cm}^{m}=0,
\end{equation}

\begin{equation}\label{28}
\alpha^{i}\beta_{i}=0,
\end{equation}

\begin{equation}\label{29}
\alpha^{n}{f}_{ni}\hspace{0cm}^{m}-\beta_{n}{\tilde{f}^{nm}}\hspace{0cm}_{i}=0,
\end{equation}

\begin{equation}\label{30}
\alpha^{i}{\tilde{f}^{mn}}\hspace{0cm}_{i}=0,
\end{equation}

\begin{equation}\label{31}
\beta_{i}{f}_{mn}\hspace{0cm}^{i}=0.
\end{equation}
Furthermore, from \eqref{14}-\eqref{17} and \eqref{19},\eqref{22} after some calculation, one can find the commutation relations between $\{X_{i}\}$ and $\{\tilde{X}^{j}\}$ as follows

\begin{equation}\label{32}
[X_i , \tilde{X}^j] =({\tilde{f}^{jk}}\hspace{0cm}_{i}+\frac{1}{2}\alpha^{k}\delta_{i}\hspace{0cm}^{j}-\alpha^{j}\delta_{i}\hspace{0cm}^{k})X_k +({f}_{ki}\hspace{0cm}^{j}-\frac{1}{2}\beta_{k}\delta_{i}\hspace{0cm}^{j}+\beta_{i}\delta_{k}\hspace{0cm}^{j}) \tilde{X}^k.
\end{equation}
Note that the relations \eqref{20}-\eqref{21} and \eqref{27}-\eqref{31} and \eqref{32} are the algebraic definitions for the Jacobi-Lie bialgebras in terms of basis  $\{X_{i}\}$ and $\{\tilde{X}^{j}\}$ and in this sense, these are a generalization of the ordinary Lie bialgebras  \eqref{19}-\eqref{21} and \eqref{24}-\eqref{25}\footnote{Note that relations \eqref{27}-\eqref{32} for $\alpha^{i}=\beta_{i}=0$, reduce to \eqref{24} and \eqref{25}.}.
\begin{definition}
A Jacobi-Lie bialgebra is a pair $(({\bf{g}},\phi_{0}),({\bf{g}}^{*},X_{0}))$ where $({\bf{g}},[,]^{{\bf{g}}})$ is a real Lie algebra of finite dimension with the Lie bracket $[,]^{\bf g}$ and the basis $\{X_{i}\}$ and Lie algebra ${\bf{g}}^{*}$ (where it is dual space of ${\bf{g}}$) with Lie bracket $[,]^{{\bf{g}^{*}}}$ and basis $\{\tilde{X}^{i}\}$, such that $X_{0}={\alpha}^{i} X_{i}\in{\bf{g}}$ and $\phi_{0}={\beta}_{j}{\tilde{X}}^{j}\in{\bf{g}}^{*}$ are 1-cocycles on ${\bf{g}}^{*}$ and ${\bf{g}}$, respectively, i.e.,

\begin{equation*}\label{19'}
[X_{i},X_{j}]={f_{ij}\hspace{0cm}}^{k} X_{k}\hspace{1mm} , \hspace{1mm}[\tilde{X}^{i},\tilde{X}^{j}]={{\tilde{f}}^{ij}\hspace{0cm}}_{k} {\tilde{X}}^{k},
\end{equation*}

\begin{equation*}\label{30'}
\alpha^{i}{\tilde{f}^{mn}}\hspace{0cm}_{i}=0~,~\beta_{i}{f}_{mn}\hspace{0cm}^{i}=0,
\end{equation*}
and we have

$$
\hspace{-.3cm}{f}_{ij}\hspace{0cm}^k{\tilde{f}^{mn}}\hspace{0cm}_{k}-{f}_{ik}\hspace{0cm}^m{\tilde{f}^{kn}}\hspace{0cm}_{j} -\\
{f}_{ik}\hspace{0cm}^n{\tilde{f}^{mk}}\hspace{0cm}_{j}-{f}_{kj}\hspace{0cm}^m{\tilde{f}^{kn}}\hspace{0cm}_{i}-\\
{f}_{kj}\hspace{0cm}^n{\tilde{f}^{mk}}\hspace{0cm}_{i}+\beta_{i}{\tilde{f}^{mn}}\hspace{0cm}_{j}-\beta_{j}{\tilde{f}^{mn}}\hspace{0cm}_{i}+\alpha^{m}{f}_{ij}\hspace{0cm}^n-\alpha^{n}{f}_{ij}\hspace{0cm}^m\\
$$
\begin{equation*}\label{27'}
+(\alpha^{k}{f}_{ik}\hspace{0cm}^m-\alpha^{m}\beta_{i})\delta_{j}\hspace{0cm}^{n}
-(\alpha^{k}{f}_{jk}\hspace{0cm}^m-\alpha^{m}\beta_{j})\delta_{i}\hspace{0cm}^{n}-(\alpha^{k}{f}_{ik}\hspace{0cm}^n-\alpha^{n}\beta_{i})\delta_{j}\hspace{0cm}^{m}
+(\alpha^{k}{f}_{jk}\hspace{0cm}^n-\alpha^{n}\beta_{j})\delta_{i}\hspace{0cm}^{m}=0,
\end{equation*}

\begin{equation*}\label{28'}
\alpha^{i}\beta_{i}=0,
\end{equation*}

\begin{equation*}\label{29'}
\alpha^{n}{f}_{ni}\hspace{0cm}^{m}-\beta_{n}{\tilde{f}^{nm}}\hspace{0cm}_{i}=0,
\end{equation*}\\
where the commutation relations between  $\{X_{i}\}$ and $\{\tilde{X}^{j}\}$ are as follows

\begin{equation*}\label{32'}
[X_i , \tilde{X}^j] =({\tilde{f}^{jk}}\hspace{0cm}_{i}+\frac{1}{2}\alpha^{k}\delta_{i}\hspace{0cm}^{j}-\alpha^{j}\delta_{i}\hspace{0cm}^{k})X_k +({f}_{ki}\hspace{0cm}^{j}-\frac{1}{2}\beta_{k}\delta_{i}\hspace{0cm}^{j}+\beta_{i}\delta_{k}\hspace{0cm}^{j}) \tilde{X}^k.
\end{equation*}
\end{definition}
These relations can be applied for finding and classifying the Jacobi-Lie bialgebras in low dimensions similar to the Lie bialgebras and Lie super bialgebras in low dimensions \cite{JR,ER}. To this aim, we prove the following proposition.
\begin{proposition}
Two Jacobi-Lie bialgebras $(({\bf{g}},\phi_{0}),({\bf{g}}^{*},X_{0}))$ and  $(({\bf{g}},{\phi}'_{0}),({{\bf{g}}^{*}}',{X}'_{0}))$ are equivalent, if there exist
$A \in Aut({\bf{g}})$ (automorphism group of the Lie algebra {\bf g}) such that
\begin{equation}\label{33}
{{{\tilde{f}^{ij}}\hspace{0cm}_n}_{(\bf {
{{\bf{g}}^{*}}'})}} =(A^{-t})^i\hspace{0cm}_{k}{{{\tilde{f}^{kl}}\hspace{0cm}_m}_{(\bf {
{\bf{g}}^{*}})}} (A^{-t})^j\hspace{0cm}_{l} (A^{t})^m\hspace{0cm}_{n},
\end{equation}
\begin{equation}\label{34}
{\alpha}'^{i}=(A^{-t})^i\hspace{0cm}_{m}{\alpha^{m}},
\end{equation}
\begin{equation}\label{35}
{\beta}'_{i}=A_{i}\hspace{0cm}^{m}{\beta_{m}},
\end{equation}
where $A_{m}\hspace{0cm}^{n}$s are the elements of the automorphism matrix $A$ for the Lie algebra $\bf{g}$ and $X'_{0}={\alpha'^{i}}X_{i}$ , $\phi'_{0}=\beta'_{j}\tilde{X}'^{j}$ so that $\{\tilde{X}'^{j}\}$ is the basis of ${{\bf{g}}^{*}}'$.
\end{proposition}
Proof: From the definition of automorphism of the Lie algebra; $A:\bf{g}\rightarrow\bf{g}$ in terms of the basis $X_{j}$ we have
\begin{equation}\label{36}
A{X}_{i}=A_{i}\hspace{0cm}^{j}{X_{j}},
\end{equation}
where $A_{i}\hspace{0cm}^{j}$s satisfy the following relation
\begin{equation}\label{37}
A_{i}\hspace{0cm}^{m} f_{mn}\hspace{0cm}^{k} A_{j}\hspace{0cm}^{n}= f_{ij}\hspace{0cm}^{l} A_{l}\hspace{0cm}^{k}.
\end{equation}
Now, applying \eqref{37}  in \eqref{27}, one can obtain relations \eqref{33}-\eqref{35} where ${\alpha}'^{i},{\beta}'_{i}$ and ${{{\tilde{f}^{ij}}\hspace{0cm}_n}_{(\bf {{{\bf{g}}^{*}}'})}}$ are satisfied in the relations \eqref{27}-\eqref{31} and these show that
$(({\bf{g}},{\phi}'_{0}),({{\bf{g}}^{*}}',{X}'_{0}))$ is also a Jacobi-Lie bialgebra and is equivalent to the Jacobi-Lie bialgebra
$(({\bf{g}},\phi_{0}),({\bf{g}}^{*},X_{0}))$.\\
The above proposition in the case of $\phi_{0}=0$ and $X_{0}=0$ recovers the equivalency between Lie bialgebras $({\bf{g}},\delta)$ and $({\bf{g}},\delta^{'})$ \cite{RHR} i.e., the relation
\begin{equation}\label{38}
\delta^{'}=(A\otimes A)\circ\delta\circ A^{-1}.
\end{equation}
In this way, one can apply the definition 4 and proposition 5 for obtaining and classifying the Jacobi-Lie bialgebras, directly.
%%%%%%%%%%%%%%%%%%%%%%%%%%%%%%%%%%%%%%%%%%%%%%%%%%%%%%%%%%%%%%%%%%%%%%%%%%%%%%%%%%%%%%%%%%%%%%
\section{\bf Calculation of Jacobi-Lie bialgebras using adjoint representation }
In this section, using adjoint representation of the Lie algebras ${\bf g}$ and ${\bf{g^{*}}}$ in \eqref{21} and \eqref{27}-\eqref{31}, we provide a procedure like \cite{ER} for calculating and classifying low dimensional Jacobi-Lie bialgebras.
Because of tensorial form of mentioned relations, working with them is very difficult, so, we suggest
writing these equations in matrix forms using the following adjoint representations for the Lie algebras ${\bf g}$ and ${\bf{g^{*}}}$

\begin{equation}\label{39}
({\cal{X}}_{i})_{j}\hspace{0cm}^{k}=-f_{ij}\hspace{0cm}^{k}~~~~~,~~~~~({\cal{Y}}^{k})_{ij}=-f_{ij}\hspace{0cm}^{k},
\end{equation}

\begin{equation}\label{40}
({\tilde{\cal{X}}}^{i})^{j}\hspace{0cm}_{k}=-{\tilde{f}}^{ij}\hspace{0cm}_{k}~~~~~,~~~~~({\tilde{\cal{Y}}}_{k})^{ij}=-{\tilde{f}}^{ij}\hspace{0cm}_{k},
\end{equation}
Then, the matrix forms of the relations \eqref{21} and \eqref{27}-\eqref{31} become as follows, respectively,

\begin{equation}\label{41}
({\tilde{\cal X}}^i)^j_{\; \;k}{\tilde{\cal X}}^k + {\tilde{\cal X}}^i
{\tilde{\cal X}}^j - {\tilde{\cal
X}}^j {\tilde{\cal X}}^i =0,
\end{equation}

\begin{equation}\label{42}
({\cal{D}}^{mn})_{ij} + {\cal{C}}_{i}\hspace{0cm}^{m}\delta_{j}\hspace{0cm}^{n} - {\cal{C}}_{j}\hspace{0cm}^{m}\delta_{i}\hspace{0cm}^{n}-
{\cal{C}}_{i}\hspace{0cm}^{n}\delta_{j}\hspace{0cm}^{m}+ {\cal{C}}_{j}\hspace{0cm}^{n}\delta_{i}\hspace{0cm}^{m}=0,
\end{equation}

\begin{equation}\label{43}
Tr({\cal{A}}{\cal{B}}^{t})=0,
\end{equation}

\begin{equation}\label{44}
{\alpha}^{i}({\cal{X}}_{i})^{t}-{\beta}_{i} {\tilde{\cal{X}}}^{i}=0,
\end{equation}

\begin{equation}\label{45}
{\alpha}^{i}{\tilde{\cal{Y}}}_{i}=0,
\end{equation}

\begin{equation}\label{46}
{\beta}_{i}{\cal{Y}}^{i}=0,
\end{equation}
where the matrices ${\cal {C}}$ and ${\cal{D}}^{mn}$ have the following forms
$$
{\cal{C}}=\alpha^{k}{\cal{X}}_{k}-{\cal{B}}{\cal{A}}^{t},
$$
\begin{equation}\label{47}
{\cal{D}}^{mn}=({\tilde{\cal{X}}}^{m})^{n}\hspace{0cm}_{k} {\cal{Y}}^{k}+{\cal{Y}}^{m}{\tilde{\cal{X}}}^{n}-{\cal{Y}}^{n}{\tilde{\cal{X}}}^{m}
+({\tilde{\cal{X}}}^{n})^{t}{\cal{Y}}^{m}-({{\tilde{\cal{X}}}^{m}})^{t}{\cal{Y}}^{n}+{\cal{B}}({\cal{\tilde F}}^{mn})^{t}-{\cal{\tilde F}}^{mn}{\cal{B}}^{t}
+\alpha^{n}{\cal{Y}}^{m}-\alpha^{m}{\cal{Y}}^{n},
\end{equation}
and $\cal{A,B}$ and ${\tilde F}$ represent the following column matrices (where $d$ is dimension of the Lie algebras ${\bf g}$ and ${\bf g}^{*}$).

\begin{equation}\label{48}
{\cal{A}}=\left(\begin{array}{c}
\alpha_{1}\\
\alpha_{2}\\
.\\
.\\
.\\
\alpha_{d}\\
\end{array}
\right),\hspace{1cm}
{\cal{B}}=\left(\begin{array}{c}
\beta_{1}\\
\beta_{2}\\
.\\
.\\
.\\
\beta_{d}\\
\end{array}
\right),\hspace{1cm}
{\cal{\tilde F}}^{mn}=\left(\begin{array}{c}
\tilde{f}^{mn}\hspace{0cm}_{1}\\
\tilde{f}^{mn}\hspace{0cm}_{2}\\
.\\
.\\
.\\
\tilde{f}^{mn}\hspace{0cm}_{d}\\
\end{array}
\right).
\end{equation}
Now, by substituting the structure constants of Lie algebra ${\bf{g}}$ in the matrix equations \eqref{42}-\eqref{46} and solving
these equations simultaneously using \eqref{41}, we obtain the structure constants of dual Lie algebras ${\bf{g^{*}}}$ and the matrices ${\cal {A,B}} $ so that $(({\bf{g}},\phi_{0}),({\bf{g}}^{*},X_{0}))$ is a Jacobi-Lie bialgebra. By this method, we will classify two and three dimensional Jacobi-Lie bialgebras.
We perform this work in the following three steps.

\bigskip

{\bf {\small Step 1:}}~~{\it Solving the equations \eqref{41}-\eqref{46} and determining the Lie algebras
$\bf g'$ which are isomorphic with dual solutions ${\bf {g^{*}}}$}

\smallskip

By solving matrix equations \eqref{41}-\eqref{46} for obtaining matrices ${\tilde{\cal X}}^i$, ${\cal{A}}$ and ${\cal{B}}$,  some structure constants of ${\bf {g^{*}}}$ and also some coefficients of $\alpha^{i}$ and $\beta_{i}$ are obtained to be zero, some are unknown and some are obtained in terms of each other. In order to know whether ${\bf {g^{*}}}$ is one of the known Lie algebras of the classification table or it is isomorphic to them, we must use the following isomorphism relation between the obtained Lie algebras ${\bf {g^{*}}}$ and one of
the known Lie algebras of the classification table, e.g. $\bf g'$. Applying the following transformation for a change of  basis ${\bf {g^{*}}}$, we have
\begin{equation}\label{49}
\tilde{X}^{'\;i}=C^i\hspace{0cm}_{j}\tilde{X}^j,\hspace{20mm}
[\tilde{X}^{'\;i} ,\tilde{X}^{'\;j}] ={\tilde{f}^{'\;ij}}\hspace{0cm}_{k} \tilde{X}^{'\;k}.
\end{equation}
Then, we obtain the following matrix equations for isomorphism
\begin{equation}\label{50}
C\;(C^i\hspace{0cm}_{k}\;\tilde{\cal X}^k_{\bf {(g^{*})}})={\cal X}^{i}_{(\bf g')}\;C,
\end{equation}
where ${\cal X}^{i}_{(\bf g')}$ are adjoint matrices
of the known Lie algebra $\bf g'$ of the classification table.
Solving equation \eqref{50} with the condition $det C\neq 0$, we obtain some
extra conditions on ${{\tilde{f}^{kl}}_{(\bf {g^{*}})}}\hspace{1mm}_m$s which were obtained from \eqref{41}-\eqref{46}.

\bigskip

{\bf {\small Step 2:}}~~{\it Obtaining general form of the
transformation matrices $B:{\bf g}'\longrightarrow {\bf g}'.i$;
such that $(({\bf g},\phi'_{0}), ({\bf g}'.i,X'_{0}))$ is a Jacobi-Lie bialgebra}

\smallskip

As the second step, we transform Jacobi-Lie bialgebra $(({\bf g},\phi_{0}), ({\bf g}^{*},X_{0}))$ (where in the Lie algebra $\bf{g}^{*}$ we impose the extra conditions obtained in the step one) to Jacobi-Lie bialgebra  $(({\bf g},\phi'_{0}), ({\bf g}'.i,X'_{0}))$ (where ${\bf g}'.i$ is isomorphic as Lie algebra with ${\bf g}'$) using an automorphism $A$ of the Lie algebra ${\bf g}$.
As the  inner product \eqref{22} is invariant, we have $A^{-t}:{\bf {g}^{*}}\longrightarrow {\bf g}'.i$,
\begin{equation}\label{51}
X'_i=A_i\hspace{0cm}^{k}
X_k,\hspace{10mm}\tilde{X}^{'\;j}=(A^{-t})^j\hspace{0cm}_{l}\tilde{X}^l,\hspace{10mm}<X'_i
, \tilde{X}^{'\;j}> = \delta_i\hspace{0cm}^{j},
\end{equation}
where $A^{-t}$ is inverse transpose of every matrix
$A\in Aut(\bf g)$. Thus, we have the following relation
\begin{equation}\label{52}
(A^{-t})^i\hspace{0cm}_{k}{{\tilde{f}^{kl}}_{(\bf {g^{*}})}}\hspace{1mm}_m (A^{-t})^j\hspace{0cm}_{l} = {{{{f}}^{ij}}_{(\bf
{g}'.i)}}\hspace{0.5mm}_n (A^{-t})^n\hspace{0cm}_{m}.\\
\end{equation}
Now, for obtaining Jacobi-Lie bialgebras $(({\bf g},\phi'_{0}), ({\bf g}'.i,X'_{0}))$,
we must obtain Lie algebras ${\bf g}'.i$ or transformations \\
$B:{\bf g}'\longrightarrow {\bf g}'.i$ such that
\begin{equation}\label{53}
B^i\hspace{0cm}_{k}{{{f}^{kl}}_{(\bf {
g}')}}\hspace{0.5mm}_m B^j\hspace{0cm}_{l} = {{{{f}}^{ij}}_
{(\bf {g}'.i)}}\hspace{0.5mm}_n B^n\hspace{0cm}_{m}.\\
\end{equation}
To this end, it is enough to omit ${{{{f}}^{ij}}_ {(\bf
{g}'.i)}}\hspace{1mm}_n $ between \eqref{52} and \eqref{53}. Then, we will have
the following matrix equation for $B$
\begin{equation}\label{54}
(A^{-t})^i\hspace{0cm}_{m}\tilde{\cal X}^{t\;m}_{\bf {(g^{*})}}A^{-1} =(B^{t} A)^{-1}(B^i\hspace{0cm}_{k}{{\cal X}^{t\;k}}_{({\bf g}')}) B^{t}.\\
\end{equation}
By solving \eqref{54}, we obtain the general form of
matrix $B$ with the condition $detB \neq 0$. In solving  equation \eqref{54}, one
can obtain conditions on the elements of the matrix $A$; and only these conditions should be considered under which we have $det A\neq 0$.

\bigskip

{\bf {\small Step 3:}}~~{\it Obtaining the
non-equivalent  Jacobi-Lie bialgebras}

\smallskip

Having solved \eqref{54}, we obtain the general form of the matrix $B$
so that its elements are written in terms of the elements of
matrices $A$, $C$ and structure constants
${{\tilde{f}^{ij}}_{(\bf {g^{*}})}}\hspace{0.5mm}_k $ and some $\alpha^{i}$ and $\beta_{i}$. Now, by
substituting $B$ in \eqref{53}, we obtain structure constants
${{{{f}}^{ij}}_ {(\bf {g}'.i)}}\hspace{0.5mm}_n $ of the Lie
algebra ${\bf g}'.i$ in terms of elements of matrices $A$ and
$C$ and some ${{\tilde{f}^{ij}}_{(\bf {g^{*}})}}\hspace{1mm}_k$s, and also using \eqref{34} and \eqref{35}, we obtain the $\alpha'^{i}$ and $\beta'_{i}$ or column matrices ${\cal A'}$ and ${\cal B'}$.
Then, we check whether it is possible to equalize some structure
constants ${{{{f}}^{ij}}_ {(\bf {g}'.i)}}\hspace{0.5mm}_n$ with
each other and with $\pm1$ such that $det A\neq0$,
$det B\neq0$ and $det C\neq0$. In this way, we obtain matrices $B_1$, $B_2$,... and also ${\cal A''}$ and ${\cal B''}$,... .
Note that in obtaining $B_i$s, we impose the condition $B{B_i}^{-1}\in
Aut^{t}(\bf g)$ (where $Aut^{t}(\bf g)$ is the transpose of
$Aut(\bf g)$); if this condition is not satisfied, we can not
impose it on the structure constants because $B$ and $B_i$ are not
equivalent.

\smallskip

Now, using isomorphism matrices $B_1$, $B_2$, ... and also ${\cal A'}$,${\cal A''}$ and ${\cal B'}$,${\cal B''}$, ... we can obtain
Jacobi-Lie bialgebras $(({\bf g},\phi'_{0}), ({\bf g}'.i,X'_{0}))$, $(({\bf g},\phi''_{0}), ({\bf g}'.ii,X''_{0}))$,... . But, there is a question: which of these Jacobi-Lie bialgebras are equivalent? To answer this
question, we use the matrix form of the relation \eqref{33}. Consider the
two Jacobi-Lie bialgebras $(({\bf g},\phi'_{0}), ({\bf g}'.i,X'_{0}))$ and $(({\bf g},\phi''_{0}), ({\bf g}'.ii,X''_{0}))$, using
\begin{equation}\label{55}
A(X_i)=A_i\hspace{0cm}^{j} X_j,
\end{equation}
the relation \eqref{33} will have the following matrix form
\begin{equation}\label{56}
A^{t}((A^{t})^i\hspace{0cm}_{k}{{\cal X}_{({\bf g}'.i)}}^k) = {{{\cal X}}_{({\bf g}'.ii)}}^i A^{t}.\\
\end{equation}
On the other side, the transformation matrix between ${\bf g}'.i$
and ${\bf g}'.ii$ is $B_2B_1^{-1}$ if $B_1:{\bf g}'\longrightarrow
{\bf g}'.i$ and $B_2:{\bf g}'\longrightarrow {\bf g}'.ii$; then we
have
\begin{equation}\label{57}
(B_2B_1^{-1}) ((B_2B_1^{-1})^i\hspace{0cm}_{k}{{\cal X}_{({\bf g}'.i)}}^k) = {{{\cal X}}_{({\bf g}'.ii)}}^i (B_2B_1^{-1}).\\
\end{equation}
A comparison of \eqref{57} with \eqref{56} reveals that if $B_2B_1^{-1}\in
A^{t}$ such that we have also ${\cal A''}=A^{-t}{\cal A'}$ and ${\cal B''}=A{\cal B'}$, then the Jacobi-Lie bialgebras $(({\bf g},\phi'_{0}), ({\bf g}'.i,X'_{0}))$ and $(({\bf g},\phi''_{0}), ({\bf g}'.ii,X''_{0}))$ are equivalent. In this way, we obtain nonequivalent classes of $B_i$s and ${\cal A}$s and ${\cal B}$s such that we consider only one element of this class. Note that, for obtaining and fixing $\alpha^{i}$, $\beta_{i}$ we must impose conditions which were obtained for the elements  ${{{{f}}^{ij}}_ {(\bf {g}'.i)}}$s and elements of automorphism group in relations \eqref{34} and \eqref{35}, then we have to fix elements of $\alpha'^{i}$ and $\beta'_{i}$ with constant values (0,1,...), used in those relations. In this manner, we obtain $\phi'_{0}$, $X'_{0}$ and then Jacobi-Lie bialgebras $(({\bf g},\phi'_{0}), ({\bf g}'.i,X'_{0}))$; such that one can classify all Jacobi-Lie bialgebras. In the next section, we apply this formulation to classify real two and three dimensional Jacobi-Lie bialgebras.

\vspace{5mm}
%%%%%%%%%%%%%%%%%%%%%%%%%%%%%%%%%%%%%%%%%%%%%%%%%%%%%%%%%%%%%%%%%%%%%%%%%%%%%%%%%%%%%%%%%%%%%
\section{\bf Classification of real two and three dimensional Jacobi-Lie bialgebras }
In this section, we will use the classification of real two and three dimensional Lie algebras and their automorphism groups. It should be noted that for real two dimensional Lie algebras, we will use the classification of the ref \cite{Patera} as table 1 and for real three dimensional case we will use
the Bianchi classification \cite{LL} as table 2.
%%%%%%%%%%%%%%%%%%%%%%%%%%%%%%%%%%%%
\begin{center}
\begin{tabular}{l l l p{15mm} }
\multicolumn{3}{c}{Table 1: \small Real two dimensional Lie algebras}\\
  \hline
  \hline
  Lie Algebra & Commutation relations \\
\hline
\hline
\vspace{2mm}
{\footnotesize$A_{1}$}& {\footnotesize $[X_{i},X_{j}]=0$}\\
\vspace{2mm}

{\footnotesize $A_{2}$} & {\footnotesize $[X_{1},X_{2}]=X_{1}$}\\
\hline
%%%%%%%%%%%%%%%%%%%%%%%%%%%%%
\end{tabular}
\end{center}
%%%%%%%%%%%%%%%%%%%%%%%%%%%%%%%%
\begin{center}
\begin{tabular}{l l l p{15mm} }
\multicolumn{3}{c}{Table 2: \small Real three dimensional Lie algebras}\\
  \hline
  \hline
  Lie Algebra & Commutation relations & Comments\\
\hline
\hline
\vspace{2mm}
{\footnotesize$I$}   &{\footnotesize $[X_{i},X_{j}]=0$}&\\
\vspace{2mm}

{\footnotesize$II$}  &\footnotesize $[X_{2},X_{3}]=X_{1}$ &\\
\vspace{2mm}

{\footnotesize$III$} &\footnotesize $[X_{1},X_{2}]=-(X_{2}+X_{3})$,\footnotesize$[X_{1},X_{3}]=-(X_{2}+X_{3})$&\\
\vspace{2mm}

{\footnotesize$IV$} &\footnotesize $[X_{1},X_{2}]=-(X_{2}-X_{3})$,\footnotesize$[X_{1},X_{3}]=-X_{3}$&\\
\vspace{2mm}

{\footnotesize$V$}  &\footnotesize $[X_{1},X_{2}]=-X_{2}$,\footnotesize$[X_{1},X_{3}]=-X_{3}$&\\
\vspace{2mm}

{\footnotesize$VI_{0}$}&\footnotesize $[X_{1},X_{3}]=X_{2}$,\footnotesize$[X_{2},X_{3}]=X_{1}$&\\
\vspace{2mm}

{\footnotesize$VI_{a}$}&\footnotesize $[X_{1},X_{2}]=-(aX_{2}+X_{3})$,\footnotesize$[X_{1},X_{3}]=-(X_{2}+aX_{3})$&{ \footnotesize $ a\in\Re-\{1\},\;\;a>0$ } \\
\vspace{2mm}

{\footnotesize$VII_{0}$}&\footnotesize $[X_{1},X_{3}]=-X_{2}$,\footnotesize$[X_{2},X_{3}]=X_{1}$&\\
\vspace{2mm}

{\footnotesize$VII_{a}$}&\footnotesize $[X_{1},X_{2}]=-(aX_{2}-X_{3})$,\footnotesize$[X_{1},X_{3}]=-(X_{2}+aX_{3})$&{ \footnotesize $a\in\Re,\;\;a>0 $ } \\
\vspace{2mm}

{\footnotesize$VIII$}&\footnotesize $[X_{1},X_{2}]=-X_{3}$,\footnotesize$[X_{1},X_{3}]=-X_{2}$,\footnotesize$[X_{2},X_{3}]=X_{1}$&\\
\vspace{2mm}

{\footnotesize$IX$}&\footnotesize $[X_{1},X_{2}]=X_{3}$,\footnotesize$[X_{1},X_{3}]=-X_{2}$,\footnotesize$[X_{2},X_{3}]=X_{1}$&\\
\hline
%%%%%%%%%%%%%%%%%%%%%%%%%%%%%
\end{tabular}
\end{center}
%%%%%%%%%%%%%%%%%%%%%%%%%%%%%%%%
As mentioned in the previous section, for obtaining Jacobi-Lie bialgebras, automorphism groups are necessary. These automorphism groups have been calculated using the transformation \eqref{37}, or in the matrix form (with condition $det A\neq 0$)\cite{RHR}(see also \cite{F}) using the following relation
\begin{equation}\label{58}
A {\cal Y}^k A^{t} = {\cal Y}^{i} A_{i}\hspace{0cm}^{k}.
\end{equation}
The results are given in table 3.
%%%%%%%%%%%%%%%%%%%%%%%%%%%%%%%%%%%%%%%%%%%%%%%%%%%%%%%%%%%%%%%%%%%%%%%%%%%%%%%%%%%%%%%%%%%%%%
\begin{center}
\begin{tabular}{l l l}
\multicolumn{3}{l}{Table 3:  \small Automorphism groups of the real two and three dimensional Lie algebras}\\
\hline
\hline
Lie Algebra & Automorphism groups & Comments \\
\hline
\vspace{2mm}
{\footnotesize $ A_{1}$}& {\footnotesize $GL(2,R)$} &\\

\vspace{2mm}
{\footnotesize $ A_{2}$}&{\footnotesize $\left(
\begin{array}{cc}
a & 0 \\
b & 1 \\
\end{array} \right)$} &~~~~~~~{\footnotesize  $a\in\Re-\{0\}$} \\

\vspace{2mm}
{\footnotesize $ I$}& {\footnotesize $GL(3,R)$} & \\

\vspace{2mm}
{\footnotesize $II$}&{\footnotesize $\left(
\begin{array}{ccc}
bf-ce & 0 & 0 \\
a & b & c \\
d & e & f
\end{array} \right)$} &~~~~~~~{\footnotesize  $a,b,c,d,e,f\in\Re$, $bf\neq ce$} \\

\vspace{2mm}
{\footnotesize $III$}&{\footnotesize $\left(
\begin{array}{ccc}
1 & a & b \\
0 & c & d \\
0 & d & c
\end{array} \right)$} &~~~~~~~{\footnotesize  $a,b,c,d\in\Re$, $c\neq\pm d$} \\

\vspace{2mm}
{\footnotesize $IV$}&{\footnotesize $\left(
\begin{array}{ccc}
1 & a & b \\
0 & c & d \\
0 & 0 & c
\end{array} \right)$} &~~~~~~~{\footnotesize  $a,b,d\in\Re$, $c\in\Re-\{0\}$} \\

\vspace{2mm}
{\footnotesize $V$}&{\footnotesize $\left(
\begin{array}{ccc}
1 & a & b \\
0 & c & d \\
0 & e & f
\end{array} \right)$} &~~~~~~~{\footnotesize  $a,b,c,d,e,f\in\Re$, $cf\neq ed$} \\

\vspace{2mm}
{\footnotesize $VI_{0}$}&{\footnotesize $\left(
\begin{array}{ccc}
a & b & 0 \\
b & a & 0 \\
c & d & 1
\end{array} \right)$},{\footnotesize $\left(
\begin{array}{ccc}
a & b & 0 \\
-b & -a & 0 \\
c & d & -1
\end{array} \right)$}&~~~~~~~{\footnotesize  $a,b,c,d\in\Re$, $a\neq\pm b$} \\

\vspace{2mm}
{\footnotesize $VI_{a}$}&{\footnotesize $\left(
\begin{array}{ccc}
1 & b & c \\
0 & d & e \\
0 & e & d
\end{array} \right)$} &~~~~~~~{\footnotesize  $b,c,d,e\in\Re$, $d\neq\pm e$} \\

\vspace{2mm}
{\footnotesize $VII_{0}$}&{\footnotesize $\left(
\begin{array}{ccc}
a & b & 0 \\
-b & a & 0 \\
c & d & 1
\end{array} \right)$},{\footnotesize $\left(
\begin{array}{ccc}
a & b & 0 \\
b & -a & 0 \\
c & d & -1
\end{array} \right)$} &~~~~~~~{\footnotesize  $a,b,c,d\in\Re$, $a^{2}+b^{2}\neq 0$} \\

\vspace{2mm}
{\footnotesize $VII_{a}$}&{\footnotesize $\left(
\begin{array}{ccc}
1 & b & c \\
0 & d & -e \\
0 & e & d
\end{array} \right)$} &~~~~~~~{\footnotesize  $b,c,d,e\in\Re$, $d^{2}+e^{2}\neq 0$} \\

\vspace{2mm}
{\footnotesize $VIII$}& {\footnotesize $SL(2,R)$}  &\\

\vspace{2mm}
{\footnotesize $IX$}& {\footnotesize $SO(3)$} &\\

\hline
\end{tabular}
\end{center}
%%%%%%%%%%%%%%%%%%%%%%%%%%%%%%%%%%%%%%%%%%%%%%%%%%%%%%%%%%%%%%%%%%%%%%%%%%%%%%%%%%%%%%%%%%%%%%
Now, using the method considered in section 3 and applying {\small MAPLE} program for solving our equations \eqref{41}-\eqref{46}, we classify
real two and three dimensional Jacobi-Lie bialgebras. Let us investigate an example for explaining the method and steps mentioned in the previous section.\\
%%%%%%%%%%%%%%%%%%%%%%%%%%%%%%%%%%%%%%%%%%%%%%%%%%%%%%%%%%%%%%%%%%%%%%%%%%%%%%%
\subsection{\bf An example}
In the following, we explain our method for this classification by describing the details of the calculations for obtaining the Jacobi-Lie bialgebra
$((III,-2\tilde{X}^{1}),(V.i,-[X_{2}+X_{3}]))$. By substituting the structure constants of Lie algebra $III$ in the matrix equations \eqref{41}-\eqref{46}, we obtain the following form for the structure constants of ${\bf{g}}^{*}$ and matrices ${\cal A,B}$
\begin{equation}\label{59}
{\tilde{f}}^{12}\hspace{0cm}_{1}={\tilde{f}}^{13}\hspace{0cm}_{1}=\alpha~,~{\tilde{f}}^{23}\hspace{0cm}_{1}=\beta~,~
{\tilde{f}}^{23}\hspace{0cm}_{2}=-{\tilde{f}}^{23}\hspace{0cm}_{3}=\gamma~,~
{\cal{A}}=\left(\begin{array}{c}
0\\
-\alpha\\
-\alpha\\
\end{array}
\right)~,~{\cal{B}}=\left(\begin{array}{c}
-2\\
0\\
0\\
\end{array}
\right).
\end{equation}
Using \eqref{50}, the obtained Lie algebra ${\bf{g}}^{*}$ is isomorphic with the Lie algebra $V$ by the following isomorphism matrix
\begin{equation}\label{60}
C=\left(
\begin{array}{ccc}
c_{11} & -\frac{\gamma c_{31}-1}{\gamma} & c_{13} \\
c_{21} & -c_{23} & c_{23} \\
c_{31} & -c_{33} & c_{33}
\end{array} \right),
\end{equation}
by assuming the conditions $\alpha=\gamma$ and $\beta=0$. Now, by substituting the above results and the automorphism group of the Lie algebra $III$ in \eqref{54} one can obtain the following form for the matrix $B$
\begin{equation}\label{61}
B=\left(
\begin{array}{ccc}
0 & b_{12} & b_{13} \\
\frac{\gamma}{c+d} & b_{22} & b_{23} \\
\frac{\gamma}{c+d} & b_{32} & b_{33}
\end{array} \right),
\end{equation}
where condition $det B\neq0$ requires $\gamma\neq 0$. Then, using \eqref{53} and \eqref{34}-\eqref{35} we have the following commutation relations for the algebra ${\bf{g'}}.i$ and the following forms for ${\cal A',B'}$
\begin{equation}\label{62}
[{\tilde X}^1,{\tilde X}^2]=\alpha'{\tilde X}^1,[{\tilde X}^1,{\tilde X}^3]=\alpha'{\tilde X}^1,[{\tilde X}^2,{\tilde X}^3]=\alpha'({\tilde X}^2-{\tilde X}^3) ,{\cal{A'}}=\left(\begin{array}{c}
0\\
-\alpha'\\
-\alpha'\\
\end{array}
\right)~,~{\cal{B'}}=\left(\begin{array}{c}
-2\\
0\\
0\\
\end{array}
\right),
\end{equation}
where $\alpha'=\frac{\gamma}{c+d}$ such that $\alpha'\neq0$ {\footnote{Here, in the above relations, $a,b,c,d$ are elements of automorphism group of the Lie algebra $III$ (see table 3).}}. Now, if $\alpha'=1$ i.e., $\gamma=c+d$ we have
\begin{equation}\label{63}
B'=\left(
\begin{array}{ccc}
0 & b'_{12} & b'_{13} \\
1 & b'_{22} & b'_{23} \\
1 & b'_{32} & b'_{33}
\end{array} \right)~,~{\cal{A''}}=\left(\begin{array}{c}
0\\
-1\\
-1\\
\end{array}
\right)~,~{\cal{B''}}=\left(\begin{array}{c}
-2\\
0\\
0\\
\end{array}
\right).
\end{equation}
Since $B'B^{-1}\in A^{t}$, then $B'$ is equivalent to $B$ and according to the relations \eqref{34} and \eqref{35}, ${\cal A'}$ and ${\cal B'}$ is equivalent to  ${\cal A''}$ and ${\cal B''}$, respectively, where $A$ is automorphism group of the Lie algebra $III$. This equivalency indicate that one can choose $\alpha'=1$. In this way, we obtain the Jacobi-Lie bialgebra $((III,\phi_{0}),(V.i,X_{0}))$
where $X_{0}=-(X_{2}+X_{3})$ and $\phi_{0}=-2\tilde{X}^{1}$ and commutation relations for $V.i$ as
\begin{equation}\label{64}
[{\tilde X}^1,{\tilde X}^2]={\tilde X}^1,[{\tilde X}^1,{\tilde X}^3]={\tilde X}^1,[{\tilde X}^2,{\tilde X}^3]={\tilde X}^2-{\tilde X}^3.
\end{equation}
Note that, we have two classes of Jacobi-Lie bialgebras; the first class is Jacobi-Lie
bialgebras with {$X_{0}$},$\phi_{0}\neq 0$ and the second class is Jacobi-Lie bialgebras with {$X_{0}$} or $ \phi_{0}=0$ {\footnote {If $(({\bf{g}},\phi_{0}),({\bf{g}}^{*},X_{0}))$ being a Jacobi-Lie bialgebra, then $(({\bf{g}}^{*},X_{0}),({\bf{g}},\phi_{0}))$ is also a Jacobi-Lie bialgebra,
therefore, in this paper (for state with {$X_{0}$} or $ \phi_{0}=0$) we have only presented Jacobi-Lie bialgebras with {$\phi_{0}=0$} .}}.
We classify real two and three dimensional Jacobi-Lie bialgebras in tables 4, 5 and 6, 7,  respectively, as follows.
\vspace{.7cm}
%%%%%%%%%%%%%%%%%%%%%%%%%%%%%%%%%%%%%%%%%%%%%%%%%%%%%%%%%%%%%%%%%%%%%%%%%%%%%
\begin{center}
\begin{tabular}{l l l l l l p{0.15mm} }
\multicolumn{6}{l}{Table 4: \small Real two dimensional Jacobi-Lie bialgebras with {\footnotesize $X_{0}$},$ \phi_{0}\neq 0$}\\
\hline
\hline
{\footnotesize ${\bf g}$ }& {\footnotesize ${\bf g^{*}}$}
&{\footnotesize Commutation relations of ${\bf g^{*}}$} &{$X_{0}$}& $\phi_{0}$&{\footnotesize Comments} \\
\hline
\vspace{2mm}
{\footnotesize $A_{1}$}&{\footnotesize $A_{1}$}&{\footnotesize $[{\tilde X}^i,{\tilde X}^j]=0$}&{\footnotesize $X_{2}$}&{\footnotesize $ {\tilde X}^{1}$}& \\
\vspace{2mm}
{\footnotesize $A_{2}$}&{\footnotesize $A_{2}.i$}&{\footnotesize $[{\tilde X}^1,{\tilde X}^2]={\tilde X}^2$}&{\footnotesize $-\alpha X_{1}$}&{\footnotesize $\alpha {\tilde X}^{2}$}& {\footnotesize $\alpha \in \Re-\{0\}$}\\
\hline
\end{tabular}
\end{center}

%%%%%%%%%%%%%%%%%%%%%%%%%%%%%%%%%%%%%%%%%%%%%%%%%%%%%%%%%%%%%%%%%%%%%%%%%%%%%%%%%%%%%%%%%%
\vspace{3mm}
\begin{center}
\begin{tabular}{l l l l l p{0.15mm} }
\multicolumn{5}{l}{Table 5: \small Real two dimensional Jacobi-Lie bialgebras with {\footnotesize $\phi_{0}=0$}}\\
\hline
\hline
{\footnotesize ${\bf g}$ }& {\footnotesize ${\bf g^{*}}$}
&{\footnotesize Commutation relations of ${\bf g^{*}}$} &{$X_{0}$}&{\footnotesize Comments} \\
\hline
\vspace{2mm}
{\footnotesize $A_{1}$}&{\footnotesize $A_{1}$}&{\footnotesize $[{\tilde X}^i,{\tilde X}^j]=0$}&{\footnotesize $X_{1}+X_{2}$}&\\
\vspace{2mm}
{\footnotesize $A_{1}$}&{\footnotesize $A_{2}$}&{\footnotesize $[{\tilde X}^1,{\tilde X}^2]={\tilde X}^1$}&{\footnotesize $\alpha X_{2}$}& {\footnotesize $\alpha \in \Re-\{0\}$}\\
\hline
\end{tabular}
\end{center}
~\\
~\\
~\\
~\\
%%%%%%%%%%%%%%%%%%%%%%%%%%%%%%%%%%%%%%%%%%%%%%%%%%%%%%%%%%%%%%%%%%%%%%%%%%%%%%%%%%%%%%%%%%
\vspace{7mm}
\hspace{-1.2cm}\begin{tabular}{ l l l l l l}
\multicolumn{6}{l}{Table 6: \small Real three dimensional Jacobi-Lie bialgebras with {\footnotesize $X_{0}$},$ \phi_{0}\neq 0$}\\
\hline
\hline
{\footnotesize ${\bf g}$ }& {\footnotesize ${\bf g^{*}}$}
&{\footnotesize Commutation relations of ${\bf g^{*}}$} &{$X_{0}$}& $\phi_{0}$&{\footnotesize Comments} \\
\hline
\vspace{2mm}
{\footnotesize $I$}&{\footnotesize $III$}&{\footnotesize $[{\tilde X}^1,{\tilde X}^2]=-({\tilde X}^2+{\tilde X}^3),[{\tilde X}^1,{\tilde X}^3]=-({\tilde X}^2+{\tilde X}^3)$}&{\footnotesize $-2X_{1}$}&{\footnotesize $-({\tilde X}^2-{\tilde X}^3)$}&\\
\vspace{2mm}
{\footnotesize $II$}&{\footnotesize $III$}&{\footnotesize $[{\tilde X}^1,{\tilde X}^2]=-({\tilde X}^2+{\tilde X}^3),[{\tilde X}^1,{\tilde X}^3]=-({\tilde X}^2+{\tilde X}^3)$}&{\footnotesize $-2X_{1}$}&{\footnotesize $-({\tilde X}^2-{\tilde X}^3)$}&\\
\vspace{2mm}
{\footnotesize $III$}&{\footnotesize $III.i$}&{\footnotesize $[{\tilde X}^2,{\tilde X}^3]=-\frac{\alpha+2}{\alpha}({\tilde X}^2+{\tilde X}^3)$}&{\footnotesize $-(X_{2}-X_{3})$}&{\footnotesize $\alpha{\tilde X}^1$}&{\footnotesize $\alpha \in \Re-\{0\}$}\\
\vspace{2mm}
{\footnotesize $III$}&{\footnotesize $III.ii$}&{\footnotesize $[{\tilde X}^1,{\tilde X}^2]={\tilde X}^1,[{\tilde X}^1,{\tilde X}^3]={\tilde X}^1$}&{\footnotesize $-2X_{2}$}&{\footnotesize $-2{\tilde X}^1$}&\\
\vspace{2mm}
{\footnotesize $III$}&{\footnotesize $III.ii$}&{\footnotesize $[{\tilde X}^1,{\tilde X}^2]={\tilde X}^1,[{\tilde X}^1,{\tilde X}^3]={\tilde X}^1$}&{\footnotesize $-(X_{2}+X_{3})$}&{\footnotesize $-2{\tilde X}^1$}&\\
\vspace{2mm}
{\footnotesize $III$}&{\footnotesize $III.iii$}&{\footnotesize $[{\tilde X}^1,{\tilde X}^2]=-{\tilde X}^1,[{\tilde X}^1,{\tilde X}^3]=-{\tilde X}^1,[{\tilde X}^2,{\tilde X}^3]=-2{\tilde X}^1$}&{\footnotesize $X_{2}+X_{3}$}&{\footnotesize $-({\tilde X}^2-{\tilde X}^3)$}&\\
\vspace{2mm}
{\footnotesize $III$}&{\footnotesize $III.iv$}&{\footnotesize $[{\tilde X}^1,{\tilde X}^2]=\frac{\alpha}{2}({\tilde X}^2+{\tilde X}^3),[{\tilde X}^1,{\tilde X}^3]=\frac{\alpha}{2}({\tilde X}^2+{\tilde X}^3),[{\tilde X}^2,{\tilde X}^3]={\tilde X}^2+{\tilde X}^3$}&{\footnotesize $\alpha X_{1}$}&{\footnotesize $-\alpha({\tilde X}^2-{\tilde X}^3)$}&{\footnotesize $\alpha \in \Re-\{0\}$}\\
\vspace{2mm}
{\footnotesize $III$}&{\footnotesize $V.i$}&{\footnotesize $[{\tilde X}^1,{\tilde X}^2]={\tilde X}^1,[{\tilde X}^1,{\tilde X}^3]={\tilde X}^1,[{\tilde X}^2,{\tilde X}^3]={\tilde X}^2-{\tilde X}^3$}&{\footnotesize $-(X_{2}+X_{3})$}&{\footnotesize $-2{\tilde X}^1$}&\\
\vspace{2mm}
{\footnotesize $IV$}&{\footnotesize $III.v$}&{\footnotesize $[{\tilde X}^1,{\tilde X}^3]={\tilde X}^1,[{\tilde X}^2,{\tilde X}^3]={\tilde X}^1$}&{\footnotesize $-X_{3}$}&{\footnotesize $-{\tilde X}^1$}&\\
\vspace{2mm}
{\footnotesize $IV$}&{\footnotesize $III.vi$}&{\footnotesize $[{\tilde X}^1,{\tilde X}^2]={\tilde X}^1,[{\tilde X}^2,{\tilde X}^3]={\tilde X}^1$}&{\footnotesize $-(X_{2}+X_{3})$}&{\footnotesize $-{\tilde X}^1$}&\\
\vspace{2mm}
{\footnotesize $IV$}&{\footnotesize $IV.i$}&{\footnotesize $[{\tilde X}^1,{\tilde X}^3]=\alpha{\tilde X}^1,[{\tilde X}^2,{\tilde X}^3]={\tilde X}^1+\alpha{\tilde X}^2$}&{\footnotesize $-\epsilon\alpha X_{3}$}&{\footnotesize $-\epsilon{\tilde X}^1$}&{\footnotesize $\epsilon=1,2~,~\alpha>0$}\\
\vspace{2mm}
{\footnotesize $IV$}&{\footnotesize $IV.ii$}&{\footnotesize $[{\tilde X}^1,{\tilde X}^3]=\alpha{\tilde X}^1,[{\tilde X}^2,{\tilde X}^3]=-{\tilde X}^1+\alpha{\tilde X}^2$}&{\footnotesize $-\epsilon\alpha X_{3}$}&{\footnotesize $-\epsilon{\tilde X}^1$}&{\footnotesize $\epsilon=1,2~,~\alpha>0$}\\
\vspace{2mm}
{\footnotesize $IV$}&{\footnotesize $V.ii$}&{\footnotesize $[{\tilde X}^1,{\tilde X}^3]={\tilde X}^1,[{\tilde X}^2,{\tilde X}^3]={\tilde X}^2$}&{\footnotesize $-\epsilon X_{3}$}&{\footnotesize $-\epsilon{\tilde X}^1$}&{\footnotesize $\epsilon=1,2$}\\
\vspace{2mm}
{\footnotesize $IV$}&{\footnotesize $VI_{0}.i$}&{\footnotesize $[{\tilde X}^1,{\tilde X}^3]={\tilde X}^1,[{\tilde X}^2,{\tilde X}^3]=-{\tilde X}^2$}&{\footnotesize $-X_{3}$}&{\footnotesize $-{\tilde X}^1$}&\\
\vspace{2mm}
{\footnotesize $IV$}&{\footnotesize $VI_{a}.i$}&{\footnotesize $[{\tilde X}^1,{\tilde X}^3]={\tilde X}^1,[{\tilde X}^2,{\tilde X}^3]=\frac{a+1}{a-1}{\tilde X}^2$}&{\footnotesize $-X_{3}$}&{\footnotesize $-{\tilde X}^1$}& {\footnotesize $a > 0~,~a\neq1$}\\
\vspace{2mm}
{\footnotesize $IV$}&{\footnotesize $VI_{a}.ii$}&{\footnotesize $[{\tilde X}^1,{\tilde X}^3]={\tilde X}^1,[{\tilde X}^2,{\tilde X}^3]=\frac{a-1}{a+1}{\tilde X}^2$}&{\footnotesize $-X_{3}$}&{\footnotesize $-{\tilde X}^1$}& {\footnotesize $a > 0~,~a\neq1$}\\
\vspace{2mm}
{\footnotesize $IV$}&{\footnotesize $VI_{a}.i$}&{\footnotesize $[{\tilde X}^1,{\tilde X}^3]={\tilde X}^1,[{\tilde X}^2,{\tilde X}^3]=\frac{a+1}{a-1}{\tilde X}^2$}&{\footnotesize $-\frac{2a}{a-1}X_{3}$}&{\footnotesize $-\frac{2a}{a-1}{\tilde X}^1$}& {\footnotesize $a > 0~,~a\neq1$}\\
\vspace{2mm}
{\footnotesize $IV$}&{\footnotesize $VI_{a}.ii$}&{\footnotesize $[{\tilde X}^1,{\tilde X}^3]={\tilde X}^1,[{\tilde X}^2,{\tilde X}^3]=\frac{a-1}{a+1}{\tilde X}^2$}&{\footnotesize $-\frac{2a}{a+1}X_{3}$}&{\footnotesize $-\frac{2a}{a+1}{\tilde X}^1$}& {\footnotesize $a > 0~,~a\neq1$}\\
\vspace{2mm}
{\footnotesize $V$}&{\footnotesize $V.i$}&{\footnotesize $[{\tilde X}^1,{\tilde X}^2]={\tilde X}^1,[{\tilde X}^1,{\tilde X}^3]={\tilde X}^1,[{\tilde X}^2,{\tilde X}^3]={\tilde X}^2-{\tilde X}^3$}&{\footnotesize $-\epsilon(X_{2}+X_{3})$}&{\footnotesize $-\epsilon{\tilde X}^1$}&{\footnotesize $\epsilon=1,2$}\\
\vspace{2mm}
{\footnotesize $V$}&{\footnotesize $VI_{0}.i$}&{\footnotesize $[{\tilde X}^1,{\tilde X}^3]={\tilde X}^1,[{\tilde X}^2,{\tilde X}^3]=-{\tilde X}^2$}&{\footnotesize $-X_{3}$}&{\footnotesize $-{\tilde X}^1$}&\\
\vspace{2mm}
{\footnotesize $V$}&{\footnotesize $VI_{a}.i$}&{\footnotesize $[{\tilde X}^1,{\tilde X}^3]={\tilde X}^1,[{\tilde X}^2,{\tilde X}^3]=\frac{a+1}{a-1}{\tilde X}^2$}&{\footnotesize $-X_{3}$}&{\footnotesize $-{\tilde X}^1$}& {\footnotesize $a > 0~,~a\neq1$}\\
\vspace{2mm}
{\footnotesize $V$}&{\footnotesize $VI_{a}.ii$}&{\footnotesize $[{\tilde X}^1,{\tilde X}^3]={\tilde X}^1,[{\tilde X}^2,{\tilde X}^3]=\frac{a-1}{a+1}{\tilde X}^2$}&{\footnotesize $-X_{3}$}&{\footnotesize $-{\tilde X}^1$}& {\footnotesize $a > 0~,~a\neq1$}\\
\vspace{2mm}
{\footnotesize $V$}&{\footnotesize $VI_{a}.i$}&{\footnotesize $[{\tilde X}^1,{\tilde X}^3]={\tilde X}^1,[{\tilde X}^2,{\tilde X}^3]=\frac{a+1}{a-1}{\tilde X}^2$}&{\footnotesize $-\frac{2a}{a-1}X_{3}$}&{\footnotesize $-\frac{2a}{a-1}{\tilde X}^1$}& {\footnotesize $a > 0~,~a\neq1$}\\
\vspace{2mm}
{\footnotesize $V$}&{\footnotesize $VI_{a}.ii$}&{\footnotesize $[{\tilde X}^1,{\tilde X}^3]={\tilde X}^1,[{\tilde X}^2,{\tilde X}^3]=\frac{a-1}{a+1}{\tilde X}^2$}&{\footnotesize $-\frac{2a}{a+1}X_{3}$}&{\footnotesize $-\frac{2a}{a+1}{\tilde X}^1$}& {\footnotesize $a > 0~,~a\neq1$}\\
\vspace{2mm}
{\footnotesize $VI_{0}$}&{\footnotesize $III.vii$}&{\footnotesize $[{\tilde X}^1,{\tilde X}^3]={\tilde X}^3,[{\tilde X}^2,{\tilde X}^3]={\tilde X}^3$}&{\footnotesize $-(X_{1}+X_{2})$}&{\footnotesize ${\tilde X}^3$}&\\
\vspace{2mm}
{\footnotesize $VI_{0}$}&{\footnotesize $III.viii$}&{\footnotesize $[{\tilde X}^1,{\tilde X}^3]=-{\tilde X}^3,[{\tilde X}^2,{\tilde X}^3]={\tilde X}^3$}&{\footnotesize $-(X_{1}-X_{2})$}&{\footnotesize ${\tilde X}^3$}&\\
\vspace{2mm}
{\footnotesize $VI_{0}$}&{\footnotesize $III.ix$}&{\footnotesize $[{\tilde X}^1,{\tilde X}^2]={\tilde X}^3,[{\tilde X}^2,{\tilde X}^3]={\tilde X}^3$}&{\footnotesize $-X_{1}$}&{\footnotesize ${\tilde X}^3$}&\\
\vspace{2mm}
{\footnotesize $VI_{0}$}&{\footnotesize $VI_{0}.ii$}&{\footnotesize $[{\tilde X}^1,{\tilde X}^2]={\tilde X}^1+{\tilde X}^2,[{\tilde X}^1,{\tilde X}^3]=-{\tilde X}^3,[{\tilde X}^2,{\tilde X}^3]={\tilde X}^3$}&{\footnotesize $-\epsilon(X_{1}-X_{2})$}&{\footnotesize $\epsilon{\tilde X}^3$}&{\footnotesize $\epsilon=1,-2$}\\
\hline
\end{tabular}
%%%%%%%%%%%%%%%%%%%%%%%%%%%%%%%%%%%%%%%%%%%%%%%%%%%%%%%%%%%%%%%%%%%%%%%%%%%%%%%%%%%%%%%%%%
\newpage
\hspace{-1.5cm}\begin{tabular}{ l l l l l l}
\multicolumn{6}{l}{Table 6: Real three dimensional Jacobi-Lie bialgebras with {\footnotesize $X_{0}$},$ \phi_{0}\neq 0$ \small(Continued.)}\\
\hline
\hline
{\footnotesize ${\bf g}$ }& {\footnotesize ${\bf g^{*}}$}
&{\footnotesize Commutation relations of ${\bf g^{*}}$} &{$X_{0}$}& $\phi_{0}$&{\footnotesize Comments} \\
\hline
\vspace{2mm}
{\footnotesize $VI_{0}$}&{\footnotesize $VI_{a}.iii$}&{\footnotesize $[{\tilde X}^1,{\tilde X}^2]=-\frac{a+1}{a-1}({\tilde X}^1+{\tilde X}^2),[{\tilde X}^1,{\tilde X}^3]=-{\tilde X}^3,[{\tilde X}^2,{\tilde X}^3]={\tilde X}^3$}&{\footnotesize $-(X_{1}-X_{2})$}&{\footnotesize ${\tilde X}^3$}& {\footnotesize $a > 0~,~a\neq1$}\\
\vspace{2mm}
{\footnotesize $VI_{0}$}&{\footnotesize $VI_{a}.iv$}&{\footnotesize $[{\tilde X}^1,{\tilde X}^2]=-\frac{a-1}{a+1}({\tilde X}^1+{\tilde X}^2),[{\tilde X}^1,{\tilde X}^3]=-{\tilde X}^3,[{\tilde X}^2,{\tilde X}^3]={\tilde X}^3$}&{\footnotesize $-(X_{1}-X_{2})$}&{\footnotesize ${\tilde X}^3$}& {\footnotesize $a > 0~,~a\neq1$}\\
\vspace{2mm}
{\footnotesize $VI_{0}$}&{\footnotesize $VI_{a}.iii$}&{\footnotesize $[{\tilde X}^1,{\tilde X}^2]=-\frac{a+1}{a-1}({\tilde X}^1+{\tilde X}^2),[{\tilde X}^1,{\tilde X}^3]=-{\tilde X}^3,[{\tilde X}^2,{\tilde X}^3]={\tilde X}^3$}&{\footnotesize $-\frac{2}{a-1}(X_{1}-X_{2})$}&{\footnotesize $\frac{2}{a-1}{\tilde X}^3$}& {\footnotesize $a > 0~,~a\neq1,3$}\\
\vspace{2mm}
{\footnotesize $VI_{0}$}&{\footnotesize $VI_{a}.iv$}&{\footnotesize $[{\tilde X}^1,{\tilde X}^2]=-\frac{a-1}{a+1}({\tilde X}^1+{\tilde X}^2),[{\tilde X}^1,{\tilde X}^3]=-{\tilde X}^3,[{\tilde X}^2,{\tilde X}^3]={\tilde X}^3$}&{\footnotesize $\frac{2}{a+1}(X_{1}-X_{2})$}&{\footnotesize $-\frac{2}{a+1}{\tilde X}^3$}& {\footnotesize $a > 0~,~a\neq1$}\\
\vspace{2mm}
{\footnotesize $VI_{a}$}&{\footnotesize $III.ii$}&{\footnotesize $[{\tilde X}^1,{\tilde X}^2]={\tilde X}^1,[{\tilde X}^1,{\tilde X}^3]={\tilde X}^1$}&{\footnotesize $-(X_{2}+X_{3})$}&{\footnotesize $-(a+1){\tilde X}^1$}& {\footnotesize $a>0~,~a\neq1$}\\
\vspace{2mm}
{\footnotesize $VI_{a}$}&{\footnotesize $III.ii$}&{\footnotesize $[{\tilde X}^1,{\tilde X}^2]={\tilde X}^1,[{\tilde X}^1,{\tilde X}^3]={\tilde X}^1$}&{\footnotesize $-\frac{a-1}{a+1}(X_{2}+X_{3})$}&{\footnotesize $-(a-1){\tilde X}^1$}& {\footnotesize $a>0~,~a\neq1$}\\
\vspace{2mm}
{\footnotesize $VI_{a}$}&{\footnotesize $III.v$}&{\footnotesize $[{\tilde X}^1,{\tilde X}^3]={\tilde X}^1,[{\tilde X}^2,{\tilde X}^3]={\tilde X}^1$}&{\footnotesize $\frac{1}{a-1}(X_{2}-aX_{3})$}&{\footnotesize $-(a+1){\tilde X}^1$}& {\footnotesize $a>0~,~a\neq1$}\\
\vspace{2mm}
{\footnotesize $VI_{a}$}&{\footnotesize $III.v$}&{\footnotesize $[{\tilde X}^1,{\tilde X}^3]={\tilde X}^1,[{\tilde X}^2,{\tilde X}^3]={\tilde X}^1$}&{\footnotesize $\frac{1}{a+1}(X_{2}-aX_{3})$}&{\footnotesize $-(a-1){\tilde X}^1$}& {\footnotesize $a>0~,~a\neq1$}\\
\vspace{2mm}
{\footnotesize $VI_{a}$}&{\footnotesize $III.x$}&{\footnotesize $[{\tilde X}^1,{\tilde X}^2]={\tilde X}^1,[{\tilde X}^1,{\tilde X}^3]=-{\tilde X}^1$}&{\footnotesize $-(X_{2}-X_{3})$}&{\footnotesize $-(a-1){\tilde X}^1$}& {\footnotesize $a>0~,~a\neq1$}\\
\vspace{2mm}
{\footnotesize $VI_{a}$}&{\footnotesize $III.x$}&{\footnotesize $[{\tilde X}^1,{\tilde X}^2]={\tilde X}^1,[{\tilde X}^1,{\tilde X}^3]=-{\tilde X}^1$}&{\footnotesize $-\frac{a+1}{a-1}(X_{2}-X_{3})$}&{\footnotesize $-(a+1){\tilde X}^1$}& {\footnotesize $a>0~,~a\neq1$}\\
{\footnotesize $VI_{a}$}&{\footnotesize $VI_{b}.v$}&{\footnotesize $[{\tilde X}^1,{\tilde X}^2]={\tilde X}^1,[{\tilde X}^1,{\tilde X}^3]={\tilde X}^1,[{\tilde X}^2,{\tilde X}^3]=\frac{b+1}{b-1}({\tilde X}^2-{\tilde X}^3)$}&{\footnotesize $-(X_{2}+X_{3})$}&{\footnotesize $-(a+1){\tilde X}^1$}& {\footnotesize $a>0~,~a\neq1$}\\
\vspace{1mm}
&&&&& {\footnotesize $b >0~,~b\neq1$}\\
{\footnotesize $VI_{a}$}&{\footnotesize $VI_{b}.vi$}&{\footnotesize $[{\tilde X}^1,{\tilde X}^2]={\tilde X}^1,[{\tilde X}^1,{\tilde X}^3]={\tilde X}^1,[{\tilde X}^2,{\tilde X}^3]=\frac{b-1}{b+1}({\tilde X}^2-{\tilde X}^3)$}&{\footnotesize $-(X_{2}+X_{3})$}&{\footnotesize $-(a+1){\tilde X}^1$}& {\footnotesize $a>0~,~a\neq1$}\\
\vspace{1mm}
&&&&& {\footnotesize $b >0~,~b\neq1$}\\
{\footnotesize $VI_{a}$}&{\footnotesize $VI_{b}.vii$}&{\footnotesize $[{\tilde X}^1,{\tilde X}^2]={\tilde X}^1,[{\tilde X}^1,{\tilde X}^3]=-{\tilde X}^1,[{\tilde X}^2,{\tilde X}^3]=-\frac{b+1}{b-1}({\tilde X}^2+{\tilde X}^3)$}&{\footnotesize $-(X_{2}-X_{3})$}&{\footnotesize $-(a-1){\tilde X}^1$}& {\footnotesize $a>0~,~a\neq1$}\\
\vspace{1mm}
&&&&& {\footnotesize $b >0~,~b\neq1$}\\
{\footnotesize $VI_{a}$}&{\footnotesize $VI_{b}.viii$}&{\footnotesize $[{\tilde X}^1,{\tilde X}^2]={\tilde X}^1,[{\tilde X}^1,{\tilde X}^3]=-{\tilde X}^1,[{\tilde X}^2,{\tilde X}^3]=-\frac{b-1}{b+1}({\tilde X}^2+{\tilde X}^3)$}&{\footnotesize $-(X_{2}-X_{3})$}&{\footnotesize $-(a-1){\tilde X}^1$}& {\footnotesize $a>0~,~a\neq1$}\\
\vspace{1mm}
&&&&& {\footnotesize $b >0~,~b\neq1$}\\
{\footnotesize $VI_{a}$}&{\footnotesize $VI_{b}.v$}&{\footnotesize $[{\tilde X}^1,{\tilde X}^2]={\tilde X}^1,[{\tilde X}^1,{\tilde X}^3]={\tilde X}^1,[{\tilde X}^2,{\tilde X}^3]=\frac{b+1}{b-1}({\tilde X}^2-{\tilde X}^3)$}&{\footnotesize $-\frac{2(ab+1)}{(a+1)(b-1)}(X_{2}+X_{3})$}&{\footnotesize $-\frac{2(ab+1)}{b-1}{\tilde X}^1$}& {\footnotesize $a>0~,~a\neq1$}\\
&&&&& {\footnotesize $b >0~,~b\neq1$}\\
\vspace{1mm}
&&&&& {\footnotesize $b \neq -\frac{a+3}{a-1}$}\\
{\footnotesize $VI_{a}$}&{\footnotesize $VI_{b}.vi$}&{\footnotesize $[{\tilde X}^1,{\tilde X}^2]={\tilde X}^1,[{\tilde X}^1,{\tilde X}^3]={\tilde X}^1,[{\tilde X}^2,{\tilde X}^3]=\frac{b-1}{b+1}({\tilde X}^2-{\tilde X}^3)$}&{\footnotesize $-\frac{2(ab-1)}{(a+1)(b+1)}(X_{2}+X_{3})$}&{\footnotesize $-\frac{2(ab-1)}{b+1}{\tilde X}^1$}& {\footnotesize $a>0~,~a\neq1$}\\
&&&&& {\footnotesize $b >0~,~b\neq1$}\\
\vspace{1mm}
&&&&& {\footnotesize $b \neq \frac{a+3}{a-1}$}\\
{\footnotesize $VI_{a}$}&{\footnotesize $VI_{b}.vii$}&{\footnotesize $[{\tilde X}^1,{\tilde X}^2]={\tilde X}^1,[{\tilde X}^1,{\tilde X}^3]=-{\tilde X}^1,[{\tilde X}^2,{\tilde X}^3]=-\frac{b+1}{b-1}({\tilde X}^2+{\tilde X}^3)$}&{\footnotesize $-\frac{2(ab-1)}{(a-1)(b-1)}(X_{2}-X_{3})$}&{\footnotesize $-\frac{2(ab-1)}{b-1}{\tilde X}^1$}& {\footnotesize $a>0~,~a\neq1$}\\
&&&&& {\footnotesize $b >0~,~b\neq1$}\\
\vspace{1mm}
&&&&& {\footnotesize $b \neq -\frac{a-3}{a+1}$}\\
{\footnotesize $VI_{a}$}&{\footnotesize $VI_{b}.viii$}&{\footnotesize $[{\tilde X}^1,{\tilde X}^2]={\tilde X}^1,[{\tilde X}^1,{\tilde X}^3]=-{\tilde X}^1,[{\tilde X}^2,{\tilde X}^3]=-\frac{b-1}{b+1}({\tilde X}^2+{\tilde X}^3)$}&{\footnotesize $-\frac{2(ab+1)}{(a-1)(b+1)}(X_{2}-X_{3})$}&{\footnotesize $-\frac{2(ab+1)}{b+1}{\tilde X}^1$}& {\footnotesize $a>0~,~a\neq1$}\\
&&&&& {\footnotesize $b >0~,~b\neq1$}\\
&&&&& {\footnotesize $b \neq \frac{a-3}{a+1}$}\\
\hline
\end{tabular}
%%%%%%%%%%%%%%%%%%%%%%%%%%%%%%%%%%%%%%%%%%%%%%%%%%%%%%%%%%%%%%%%%%%%%%%%%%%%%%%%%%%%%
\newpage
\begin{center}
\begin{tabular}{l l l l lp{0.15mm} }
\multicolumn{5}{l}{Table 7: \small Real three dimensional Jacobi-Lie bialgebras with {\footnotesize $\phi_{0}=0$}}\\
\hline
\hline
{\footnotesize ${\bf g}$ }& {\footnotesize ${\bf g^{*}}$}
&{\footnotesize Commutation relations of ${\bf g^{*}}$}&{$X_{0}$}&{\footnotesize Comments} \\
\hline
\vspace{2mm}
{\footnotesize $I$}&{\footnotesize $I$}&{\footnotesize $[{\tilde X}^i,{\tilde X}^j]=0$}&{\footnotesize $X_{1}$}&\\
\vspace{2mm}
{\footnotesize $I$}&{\footnotesize $II$}&{\footnotesize $[{\tilde X}^2,{\tilde X}^3]={\tilde X}^1$}&{\footnotesize $X_{3}$}&\\
\vspace{2mm}
{\footnotesize $I$}&{\footnotesize $III$}&{\footnotesize $[{\tilde X}^1,{\tilde X}^2]=-({\tilde X}^2+{\tilde X}^3),[{\tilde X}^1,{\tilde X}^3]=-({\tilde X}^2+{\tilde X}^3)$}& {\footnotesize $bX_{1}$}&{\footnotesize $b\in \Re-\{0\}$}\\
\vspace{2mm}
{\footnotesize $I$}&{\footnotesize $III$}&{\footnotesize $[{\tilde X}^1,{\tilde X}^2]=-({\tilde X}^2+{\tilde X}^3),[{\tilde X}^1,{\tilde X}^3]=-({\tilde X}^2+{\tilde X}^3)$}& {\footnotesize $-(X_{2}-X_{3})$}&\\
\vspace{2mm}
{\footnotesize $I$}&{\footnotesize $IV$}&{\footnotesize $[{\tilde X}^1,{\tilde X}^2]=-({\tilde X}^2-{\tilde X}^3),[{\tilde X}^1,{\tilde X}^3]=-{\tilde X}^3$}& {\footnotesize $bX_{1}$}&{\footnotesize $b\in \Re-\{0\}$}\\
\vspace{2mm}
{\footnotesize $I$}&{\footnotesize $V$}&{\footnotesize $[{\tilde X}^1,{\tilde X}^2]=-{\tilde X}^2,[{\tilde X}^1,{\tilde X}^3]=-{\tilde X}^3$}& {\footnotesize $bX_{1}$}&{\footnotesize $b\in \Re-\{0\}$}\\
\vspace{2mm}
{\footnotesize $I$}&{\footnotesize $VI_{0}$}&{\footnotesize $[{\tilde X}^1,{\tilde X}^3]={\tilde X}^2,[{\tilde X}^2,{\tilde X}^3]={\tilde X}^1$}& {\footnotesize $bX_{3}$}&{\footnotesize $b>0$}\\
{\footnotesize $I$}&{\footnotesize $VI_{a}$}&{\footnotesize $[{\tilde X}^1,{\tilde X}^2]=-(a{\tilde X}^2+{\tilde X}^3),[{\tilde X}^1,{\tilde X}^3]=-({\tilde X}^2+a{\tilde X}^3)$}& {\footnotesize $bX_{1}$}&{\footnotesize $a>0,a\neq1$}\\
&&&&{\footnotesize $b\in \Re-\{0\}$}\\
\vspace{2mm}
{\footnotesize $I$}&{\footnotesize $VII_{0}$}&{\footnotesize $[{\tilde X}^1,{\tilde X}^3]=-{\tilde X}^2,[{\tilde X}^2,{\tilde X}^3]={\tilde X}^1$}& {\footnotesize $bX_{3}$}&{\footnotesize $b>0$}\\
{\footnotesize $I$}&{\footnotesize $VII_{a}$}&{\footnotesize $[{\tilde X}^1,{\tilde X}^2]=-(a{\tilde X}^2-{\tilde X}^3),[{\tilde X}^1,{\tilde X}^3]=-({\tilde X}^2+a{\tilde X}^3)$}& {\footnotesize $bX_{1}$}&{\footnotesize $a>0$}\\
&&&&{\footnotesize $b\in \Re-\{0\}$}\\
\vspace{2mm}
{\footnotesize $II$}&{\footnotesize $I$}&{\footnotesize $[{\tilde X}^i,{\tilde X}^j]=0$}&{\footnotesize $X_{1}$}&\\
\vspace{2mm}
{\footnotesize $II$}&{\footnotesize $II.i$}&{\footnotesize $[{\tilde X}^1,{\tilde X}^3]={\tilde X}^2$}&{\footnotesize $X_{1}$}&\\
\vspace{2mm}
{\footnotesize $II$}&{\footnotesize $II.ii$}&{\footnotesize $[{\tilde X}^1,{\tilde X}^3]=-{\tilde X}^2$}&{\footnotesize $X_{1}$}&\\
\vspace{2mm}
{\footnotesize $II$}&{\footnotesize $III$}&{\footnotesize $[{\tilde X}^1,{\tilde X}^2]=-({\tilde X}^2+{\tilde X}^3),[{\tilde X}^1,{\tilde X}^3]=-({\tilde X}^2+{\tilde X}^3)$}&{\footnotesize $b X_{1}$}&{\footnotesize $b\in \Re-\{0\}$}\\
\vspace{2mm}
{\footnotesize $II$}&{\footnotesize $IV$}&{\footnotesize $[{\tilde X}^1,{\tilde X}^2]=-({\tilde X}^2-{\tilde X}^3),[{\tilde X}^1,{\tilde X}^3]=-{\tilde X}^3$}&{\footnotesize $b X_{1}$}&{\footnotesize $b\in \Re-\{0\}$}\\
\vspace{2mm}
{\footnotesize $II$}&{\footnotesize $IV.iii$}&{\footnotesize $[{\tilde X}^1,{\tilde X}^2]={\tilde X}^2-{\tilde X}^3,[{\tilde X}^1,{\tilde X}^3]={\tilde X}^3$}&{\footnotesize $b X_{1}$}&{\footnotesize $b\in \Re-\{0\}$}\\
\vspace{2mm}
{\footnotesize $II$}&{\footnotesize $V$}&{\footnotesize $[{\tilde X}^1,{\tilde X}^2]=-{\tilde X}^2,[{\tilde X}^1,{\tilde X}^3]=-{\tilde X}^3$}&{\footnotesize $b X_{1}$}&{\footnotesize $b\in \Re-\{0\}$}\\
\vspace{2mm}
{\footnotesize $II$}&{\footnotesize $VI_{0}.iii$}&{\footnotesize $[{\tilde X}^1,{\tilde X}^2]={\tilde X}^3,[{\tilde X}^1,{\tilde X}^3]={\tilde X}^2$}&{\footnotesize $b X_{1}$}&{\footnotesize $b >0$}\\
{\footnotesize $II$}&{\footnotesize $VI_{a}$}&{\footnotesize $[{\tilde X}^1,{\tilde X}^2]=-(a{\tilde X}^2+{\tilde X}^3),[{\tilde X}^1,{\tilde X}^3]=-({\tilde X}^2+a{\tilde X}^3)$}&{\footnotesize $b {X}_1$}& {\footnotesize $a>0,a\neq1$}\\
&&&&{\footnotesize $b \in \Re-\{0\}$}\\
\vspace{2mm}
{\footnotesize $II$}&{\footnotesize $VII_{0}.i$}&{\footnotesize $[{\tilde X}^1,{\tilde X}^2]=-{\tilde X}^3,[{\tilde X}^1,{\tilde X}^3]={\tilde X}^2$}& {\footnotesize $bX_{1}$}&{\footnotesize $b>0$}\\
\vspace{2mm}
{\footnotesize $II$}&{\footnotesize $VII_{0}.ii$}&{\footnotesize $[{\tilde X}^1,{\tilde X}^2]={\tilde X}^3,[{\tilde X}^1,{\tilde X}^3]=-{\tilde X}^2$}& {\footnotesize $bX_{1}$}&{\footnotesize $b>0$}\\
{\footnotesize $II$}&{\footnotesize $VII_{a}$}&{\footnotesize $[{\tilde X}^1,{\tilde X}^2]=-(a{\tilde X}^2-{\tilde X}^3),[{\tilde X}^1,{\tilde X}^3]=-({\tilde X}^2+a{\tilde X}^3)$}&{\footnotesize $b {X}_1$}& {\footnotesize $a>0$}\\
&&&&{\footnotesize $b \in \Re-\{0\}$}\\
{\footnotesize $II$}&{\footnotesize $VII_{a}.i$}&{\footnotesize $[{\tilde X}^1,{\tilde X}^2]=a{\tilde X}^2-{\tilde X}^3,[{\tilde X}^1,{\tilde X}^3]={\tilde X}^2+a{\tilde X}^3$}&{\footnotesize $b {X}_1$}& {\footnotesize $a>0$}\\
&&&&{\footnotesize $b \in \Re-\{0\}$}\\
\vspace{2mm}
{\footnotesize $III$}&{\footnotesize $III.v$}&{\footnotesize $[{\tilde X}^1,{\tilde X}^3]={\tilde X}^1,[{\tilde X}^2,{\tilde X}^3]={\tilde X}^1$}&{\footnotesize $\frac{1}{2}(X_{2}-X_{3})$}&\\
\vspace{2mm}
{\footnotesize $III$}&{\footnotesize $III.x$}&{\footnotesize $[{\tilde X}^1,{\tilde X}^2]={\tilde X}^1,[{\tilde X}^1,{\tilde X}^3]=-{\tilde X}^1$}&{\footnotesize $-(X_{2}-X_{3})$}&\\
\vspace{2mm}
{\footnotesize $III$}&{\footnotesize $IV.iv$}&{\footnotesize $[{\tilde X}^1,{\tilde X}^2]=-{\tilde X}^1,[{\tilde X}^1,{\tilde X}^3]={\tilde X}^1,[{\tilde X}^2,{\tilde X}^3]={\tilde X}^1+{\tilde X}^2+{\tilde X}^3$}&{\footnotesize $X_{2}-X_{3}$}&\\
\vspace{2mm}
{\footnotesize $III$}&{\footnotesize $V.iii$}&{\footnotesize $[{\tilde X}^1,{\tilde X}^2]=-{\tilde X}^1,[{\tilde X}^1,{\tilde X}^3]={\tilde X}^1,[{\tilde X}^2,{\tilde X}^3]={\tilde X}^2+{\tilde X}^3$}&{\footnotesize $X_{2}-X_{3}$}&\\
\vspace{2mm}
{\footnotesize $III$}&{\footnotesize $VI_{0}.iv$}&{\footnotesize $[{\tilde X}^1,{\tilde X}^2]=-{\tilde X}^1,[{\tilde X}^1,{\tilde X}^3]={\tilde X}^1,[{\tilde X}^2,{\tilde X}^3]=-({\tilde X}^2+{\tilde X}^3)$}&{\footnotesize $X_{2}-X_{3}$}&\\
\vspace{2mm}
{\footnotesize $III$}&{\footnotesize $VI_{a}.vii$}&{\footnotesize $[{\tilde X}^1,{\tilde X}^2]={\tilde X}^1,[{\tilde X}^1,{\tilde X}^3]=-{\tilde X}^1,[{\tilde X}^2,{\tilde X}^3]=-\frac{a+1}{a-1}({\tilde X}^2+{\tilde X}^3)$}&{\footnotesize $-(X_{2}-X_{3})$}& {\footnotesize $a>0,a\neq1$}\\
\vspace{2mm}
{\footnotesize $III$}&{\footnotesize $VI_{a}.viii$}&{\footnotesize $[{\tilde X}^1,{\tilde X}^2]={\tilde X}^1,[{\tilde X}^1,{\tilde X}^3]=-{\tilde X}^1,[{\tilde X}^2,{\tilde X}^3]=-\frac{a-1}{a+1}({\tilde X}^2+{\tilde X}^3)$}&{\footnotesize $-(X_{2}-X_{3})$}& {\footnotesize $a>0,a\neq1$}\\
\hline
\end{tabular}
\end{center}

%%%%%%%%%%%%%%%%%%%%%%%%%%%%%%%%%%%%%%%%%%%%%%%%%%%%%%%%%%%%%%%%%%%%%%%%%%%%%%%%%%%%%%%%%%
\section{\bf Conclusion}
In this paper, we have described the definition of the Jacobi (generalized)-Lie bialgebras $(({\bf{g}},\phi_{0}),({\bf{g}}^{*},X_{0}))$ in terms of structure constants of the Lie algebras ${\bf g}$ and ${\bf g^{*}}$ and components of their 1-cocycles $X_{0}\in {\bf{g}}$ and $\phi_{0}\in {\bf{g}}^{*}$. In this way, we have obtained a method to classify real low dimensional Jacobi-Lie bialgebras. By this method, we have classified real two and three dimensional Jacobi-Lie bialgebras. Now, using generalized coboundary equation presented in \cite{Iglesias}, one can obtain the classical $r$-matrices of these Jacobi-Lie bialgebras and Jacobi brackets for their Jacobi-Lie groups. There are some physical applications in this direction; such as constructing integrable models, quantizing these Jacobi-Lie bialgebras, generalizing Poisson-Lie symmetry \cite{KL} to Jacobi-Lie symmetry and so on. Some of these problems are under investigation \cite{RS2},\cite{RS1}.\\

%%%%%%%%%%%%%%%%%%%%%%%%%%%%%%%%%%%%%%%%%%%%%%%%%%%%%%%%%%%%%%%%%%%%%%%%%%%%%%%%%%%%%%%%%%%
{\bf Acknowledgments}\\
This research was supported by a research fund No. 217/D/1639 from Azarbaijan Shahid Madani University.
The authors are grateful to A. Basaki for his useful aids. Also, the authors would like to thank F. Darabi, A. Eghbali and R.Gholizadeh-Roshanagh for their valuable comments and carefully reading the manuscript.
\vspace{-4mm}
%%%%%%%%%%%%%%%%%%%%%%%%%%%%%%%%%%%%%%%%%%%%%%%%%%%%%%%%%%%%%%%%%%%%%%%%%%%%%%%%%%%%%%%%%%%%%

\end{document}